\documentclass{aa}

\usepackage{graphicx}
\usepackage{booktabs}
\usepackage{amsmath}
\usepackage{url}
\usepackage{array}
\usepackage{longtable}
\usepackage{tabularx}
\usepackage{float}
\usepackage{lineno}
\bibpunct{(}{)}{;}{a}{}{,}

\newcommand{\Porb}{P_{\rm orb}}
\newcommand{\Plong}{P_{\rm long}}
\newcommand{\OC}{O\!-\!C}
\newcommand{\phL}{\Phi_{\rm L}}
\newcommand{\Rt}{R_{\rm tidal}}

\msnr{aa62548-26}
\date{Accepted 30 October 2026}

\begin{document}

\title{Photometric eclipse-center variations over the long cycles of Magellanic double-periodic variables}
\titlerunning{Eclipse-center variations in Magellanic DPVs}
\author{R. E. Mennickent\inst{1}}
\authorrunning{Mennickent}
\institute{Departamento de Astronom{\'i}a, Universidad de Concepci{\'o}n,
Casilla 160-C, Concepci{\'o}n, Chile}

\abstract
{Double-periodic variables (DPVs) are interacting binaries with an orbital modulation and a longer cycle of uncertain origin. Their photometric eclipse centers can change without a change in stellar conjunction time.}
{I searched for long-cycle-dependent eclipse-center variations in Magellanic DPVs and tested whether changing disk brightness can produce analogous shifts with fixed conjunctions.}
{I analyzed Optical Gravitational Lensing Experiment (OGLE) photometry for 109 eclipsing-DPV candidates, revised uncertain periods, and visually curated 55 systems. In separate long-phase intervals, I measured centers of eclipses by profile fitting to local flux minima ($Q$) and by shifting a fixed eclipse template ($T$). Conjunction-level resampling and false-discovery-rate control tested their long-cycle dependence; eclipse shape and matched $V-I$ colors supplied complementary diagnostics. A single synthetic semidetached DPV provided a separate numerical experiment.}
{Of 49 curated systems with sufficient timing data, 27 show at least one $Q$ or $T$ modulation under the adopted procedure. Some selections depend on the fitting choice, but ten systems retain a selected $Q$ curve with all tested local functions and windows. Three have both estimators selected in both eclipses, without establishing whether conjunction times change. Using local color interpolation, 14 of 16 systems measured in both eclipses are redder at the deeper, phase-0 eclipse. In the artificial system, fixed conjunctions coexist with large photometric shifts and quadrature-brightness changes, although the large tilted-disk cases do not reproduce the selected observed crossing phases.}
{Long-cycle-dependent photometric eclipse-center variations occur in a substantial fraction of the evaluable, curated sample, not necessarily of all DPVs. Circumstellar brightness changes can produce such shifts, but the single artificial system neither fits an observed DPV nor establishes disk tilt as the cause. Genuine conjunction-time changes remain possible in individual objects.}

\maketitle
\nolinenumbers

\section{Introduction}\label{sec:introduction}

Roche-lobe overflow changes the masses, angular momentum, and subsequent evolution of close binaries. During an active transfer phase, the observer sees not only two stars but also material in the stream and, often, an accretion disk. Their changing contributions to the light can reveal the transfer geometry, although they can also complicate the interpretation of eclipses \citep{MarchantBodensteiner2024,Mennickent2016}.

Double-periodic variables (DPVs) are semidetached binaries with an orbital modulation and a second photometric cycle usually tens of times longer than the orbit \citep{Mennickent2003,Poleski2010,Mennickent2016}. Its origin remains uncertain. Proposed explanations include changes in mass transfer or outflow, magnetic regulation associated with the donor, and changes in circumstellar geometry \citep{Mennickent2016,Schleicher2017,Jiao2026}. In one geometrical picture, a disk inclined to the orbital plane undergoes nodal precession, changing its projected area and occultation pattern \citep{Papaloizou1995,Jiao2026}. The existence or maintenance of such a tilt is a separate physical question from its observable photometric consequences.

An eclipse probes the projected light distribution more selectively than the brightness outside eclipse. Its minimum is therefore a photometric center, not necessarily the time at which the stellar centers are in conjunction. For example, a varying disk rim or stream-impact region can alter one wing of an eclipse and move its minimum without changing the orbit. Conversely, a real change in conjunction time could shift both eclipses. Distinguishing these possibilities requires attention to both eclipses, to the definition of the measured center, and to concurrent changes in eclipse shape; timing alone cannot provide a unique physical identification.

Long-cycle-dependent eclipse shifts reported for OGLE-LMC-DPV-062, together with changes in its inferred disk brightness distribution, motivated a wider search \citep{Mennickent2026DPV062}. That object is included here as OGLE-LMC-ECL-12848. A study of the Galactic-bulge DPV OGLE-BLG-ECL-275974, submitted, provides a further motivation but does not enter the present Magellanic sample. The long OGLE-III and OGLE-IV time series make a systematic search possible, although irregular cadence, period aliases, and unequal sampling of the two eclipses make an automatic detection list inadequate on its own.

This paper establishes which curated Magellanic DPVs show long-cycle-dependent {photometric} eclipse-center shifts. I first explain how the light curves and periods led to the final sample, then how individual measurements contribute to long-phase eclipse profiles and their uncertainties. I examine the dependence of the result on the center-measurement procedure before interpreting the observed timing, morphology, and color. Finally, I use one artificial semidetached system to ask whether changing circumstellar brightness can produce similar kinds of observables while conjunction is fixed. That calculation explores a possibility; it is not a fit to an OGLE light curve or a survey of DPV parameter space.

\section{OGLE data and sample construction}\label{sec:data}

\subsection{Source photometry and period review}

The input consists of candidates classified as DPVs in the Large and Small Magellanic Clouds and cross-matched to eclipsing light curves in the OGLE Collection of Variable Stars (OCVS); it is {not} the catalog of all Magellanic eclipsing binaries \citep{Poleski2010,Udalski2015,Pawlak2016,Glowacki2024}. The public catalog and time series are publicly available\footnote{\url{https://ogle.astrouw.edu.pl/cont/4_main/var/ogleiv/ecl_ms/}}. Restricting the reconciled inventory to ECL light-curve files gives 109 candidates, 74 in the LMC and 35 in the SMC. The automatic inventory contains 295\,409 $I$ measurements, with 512--7156 points per object and baselines of 7.60--27.41 yr; 20\,523 $V$ measurements serve only for colors and photometric controls. OGLE-III/IV membership was reconstructed by observing campaign. A campaign zero-point offset was allowed only where supported for an individual object, not imposed across the sample.

Orbital-period candidates were sought with box least squares and phase-dispersion minimization algorithms (PDM; \citealt{Kovacs2002,Stellingwerf1978}); long periods were examined with PDM and generalized Lomb--Scargle periodograms \citep{Zechmeister2009}. I compared aliases and period factors using phase coverage, model comparison, resampling stability, and the published OGLE clocks as external checks. The periods used below are the {adopted periods of this analysis}, not a transcription of catalog values. Their parenthetical errors in Table~\ref{tab:referee_selected55} come from delete-one-block jackknife refits: one observing season at a time for $\Porb$ and one long-cycle block at a time for $\Plong$. They describe local sensitivity within the adopted alias family; distinct aliases remain a separate caution, not an error included in parentheses.

\subsection{From automatic candidates to a curated sample}

The first, uniform automatic pass provided period candidates, coverage tests, timing measurements, and an atlas for all 109 objects. Visual examination of this atlas retained 29 systems (branch V1). A further 47 were sent to directed period revision; after replacing suspect orbital or long periods and repeating the downstream analysis, visual examination retained 26 more (branch V2). Thus, the final 55 are the disjoint union V1+V2, as shown in Table~\ref{tab:referee_selection_flow}. In particular, an object rejected by an early automatic test could be accepted after a corrected period produced a credible folded light curve and usable eclipses. The 33 objects not sent to revision and the 21 revised but not retained should not be interpreted as systems without eclipse-timing shifts.

\begin{table}
\caption{Selection and timing-evaluability flow.}
\label{tab:referee_selection_flow}
\centering\small
\begin{tabular}{@{}lr@{}}
\hline\hline
Stage or disjoint branch & Systems \\
\hline
Reconciled ECL-file candidate inventory & 109 \\
Retained after first visual atlas (V1) & 29 \\
Sent to targeted period revision & 47 \\
Retained after that revision (V2) & 26 \\
Not sent to revision and not retained & 33 \\
Revised but not retained & 21 \\
\hline
Final curated union (V1+V2) & 55 \\
Timing-evaluable among the 55 & 49 \\
At least one $Q$ or $T$ signal selected among the 49 & 27 \\
\hline
\end{tabular}
\tablefoot{The first-atlas output is not an automatic final classification. The 54 entries outside V1+V2 are not timing non-detections; their present evaluability is audited in Table~\ref{tab:referee_nonretained54}.}
\end{table}

The subsequent timing analysis requires a secure working ephemeris, sufficient measurements on both sides of an eclipse in enough intervals of long-cycle phase, and at least one valid center-measurement procedure. Forty-nine of the 55 meet these requirements for at least one eclipse and estimator; six remain in the curated catalog but cannot be tested at the required long-phase resolution. Adequate coverage of {one} eclipse is useful and was not rejected solely because the other eclipse was poorly sampled. Table~\ref{tab:referee_selected55} reports the contributing photometry and usable measurements for each curated object.

Table~\ref{tab:referee_nonretained54} evaluates the other 54 candidates under the present timing rules and gives a reason for each object's present ineligibility. This is a {retrospective} assessment, not a reconstructed log of every original visual decision. Five have no secure long clock, 48 lack the required long-phase eclipse coverage, and one retains only a marginal eclipse. The visual re-examination left OGLE-LMC-ECL-02547 ambiguous and adopted 206.285 and 188.762 d as the long periods of OGLE-LMC-ECL-34137 and 35461, respectively; those revisions did not make either object timing-evaluable. Visual curation and OGLE cadence both affect the sample. Without a measured probability of entry and detection from a defined parent DPV population, fractions below cannot be interpreted as completeness-corrected occurrence rates.

\section{Observational measurements}\label{sec:methods}

\subsection{Fixed ephemerides, long phase, and eclipse profiles}

To compare eclipse centers without allowing the measured shifts to redefine the clock, I fixed a linear circular-orbit ephemeris for each object,
\begin{equation}
C_0(E)=T_0+\Porb E,\qquad C_{0.5}(E)=T_0+\Porb(E+1/2).
\label{eq:ephemeris}
\end{equation}
Here $E$ is the integer orbital-cycle number. The fixed epoch $T_0$ defines orbital phase zero at the deeper mean $I$-band eclipse; phase 0.5 denotes the other eclipse. These labels do not identify which star is in front. The circular ephemeris is a reference convention, not a measurement of eccentricity. Long phase is $\phL=\operatorname{frac}[(t-T_{\rm L,max})/\Plong]$, with phase zero at maximum long-cycle brightness. Table~\ref{tab:referee_selected55} gives both adopted periods and epochs. The epochs are fixed origins of the adopted phase profiles, not timings of individual eclipses or long-cycle maxima. No long-phase-dependent timing shift was applied to the ephemeris.

Because individual OGLE eclipses are usually too sparsely sampled to time separately, I combined points from distinct orbital cycles in {non-overlapping intervals of long phase}. Each valid $Q$ or $T$ value, defined later in this section, is the center of one such composite eclipse profile, not the measured time of a single conjunction. There was no separate visual acceptance of individual eclipse events: finite $I$ measurements with positive errors entered according to orbital phase and long-phase interval, and the coverage and fit tests acted on the resulting composite profile. The analysis used 20 long-phase intervals when both eclipses supported at least 14 valid profiles; otherwise it used ten and required at least seven valid profiles for the eclipse examined. A profile required bilateral coverage, at least three contributing orbital cycles, and the estimator-specific validity checks. One calculated conjunction with a point in the fitting region is counted as contributing, but this does not imply that a complete eclipse was observed. Table~\ref{tab:referee_selected55} consequently distinguishes the number of $I$ points, distinct contributing conjunctions, covered long-phase intervals, and valid $Q$/$T$ measurements for each operational eclipse.

\subsection{Two definitions of a photometric center}\label{sec:qt}

The flux minimum offers a direct center measurement, but no single local function can describe every asymmetric disk eclipse. I express each measured magnitude relative to the median magnitude of the points in the fitted window, giving the dimensionless relative flux $F=10^{-0.4(m-m_{\rm med})}$. Near a smooth, non-degenerate minimum at $x_0$, where $F'(x_0)=0$ and $F''(x_0)>0$, a Taylor expansion gives $F(x)=F(x_0)+\frac{1}{2}F''(x_0)(x-x_0)^2+O[(x-x_0)^3]$. I therefore fitted the lowest-order local function with an interior minimum:
\begin{equation}
 F(x)=a_Qx^2+b_Qx+c_Q,\qquad a_Q>0,\qquad x_Q=-\frac{b_Q}{2a_Q}.
 \label{eq:q}
\end{equation}
Here $x$ is orbital phase measured from the calculated eclipse center, so $x=0$ at the fixed ephemeris. The true photometric minimum $x_0$ need not be zero; $x_Q$ estimates it from the data. The parabola approximates only the minimum region, not the entire eclipse. Fits use the propagated flux errors and reduce the influence of large residuals. Starting at the calculated center, the fitting window is updated around the fitted vertex for at most four iterations; the ephemeris itself remains unchanged. Nominal half-widths are 0.06 orbital cycles at phase 0 and 0.08 at phase 0.5. I also tested 0.75 and 1.25 times these widths. If the nominal window failed, the nearest valid alternative was used. Each wing required at least five points, with positive curvature and a vertex inside the observed interval and the nominal bounds. Forty of 1196 valid baseline $Q$ profiles needed an alternative width. The window and cubic-function checks are reported in Sect.~\ref{sec:sensitivity}.

A second center definition, T, measures the shift that best aligns the {full reference profile} with the observed eclipse. For each system and eclipse separately, I formed one reference profile $P_{\rm ref}(x)$ from all finite $I$ measurements within twice the nominal half-width about the calculated center, before grouping by long phase. Magnitudes were median-binned onto 241 orbital-phase cells, empty cells were interpolated, and the resulting profile was Gaussian-smoothed with a width of two cells. This global reference was held fixed for every long-phase interval and for the baseline resampling; no profile was recentered on its measured minimum. For each composite profile, the T fit in magnitudes is
\begin{equation}
 m(x)=c+A_T\!\left[P_{\rm ref}(x-\delta_T)-\widetilde P_{\rm ref}\right],
 \label{eq:t}
\end{equation}
where $c$ adjusts the vertical level, $A_T$ the eclipse scale, $\delta_T$ the translation, and $\widetilde P_{\rm ref}$ is the constant median magnitude of the reference. The phase coordinate is tied to Eq.~\eqref{eq:ephemeris}, but $\delta_T=0$ means no translation of the reference, not necessarily a minimum at $x=0$. If its minimum is at $x_{\rm min,ref}$, the translated minimum is at $x_{\rm min,ref}+\delta_T$. This constant reference offset does not alter the long-cycle harmonic. The two measurements are in orbital cycles; $\Delta t=\Delta\phi\,P_{\rm orb}\,(1440\ {\rm min\,d^{-1}})$ converts a displacement to minutes using that system's period.

\begin{figure*}
\centering
\includegraphics[width=\textwidth]{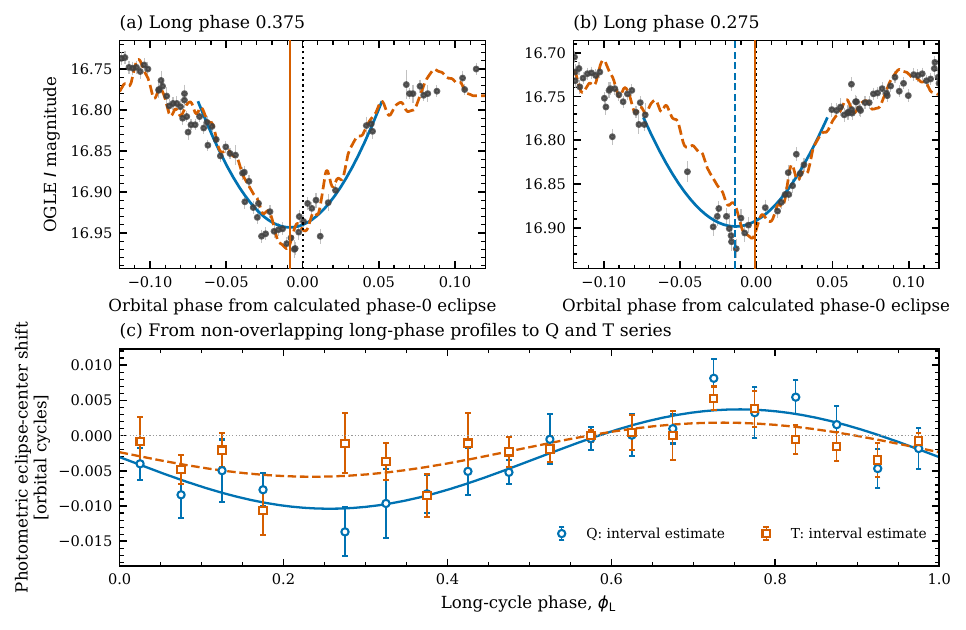}
\caption{From OGLE photometry to photometric eclipse-center variations for the phase-0 eclipse of OGLE-LMC-ECL-05935. Panels (a) and (b) contain composite profiles in two non-overlapping long-phase intervals. Points are $I$ measurements, blue curves are local $Q$ fits represented in magnitudes, and orange curves are translated fixed-template $T$ fits. The black dotted line marks the calculated eclipse center. Blue vertical lines mark $Q$ minima; orange lines mark the template translation $\delta_T$, not the absolute minimum of an asymmetric template. In panel (b), $\delta_T$ lies close to zero on the scale shown. 
{In panel (a) \(N_{\mathrm{phot}}=73\),
\(N_{\mathrm{conj},Q}=13\).
In panel (b) \(N_{\mathrm{phot}}=81\),
\(N_{\mathrm{conj},Q}=12\).
}
$N_{\rm phot}$ refers to points within $\pm0.12$ orbital cycles and $N_{\rm conj,Q}$ to distinct calculated conjunctions that supply a point to the local $Q$ window, not to complete light curves. Panel (c) shows the resulting $Q$ and $T$ displacements, bootstrap standard errors, and first-harmonic fits versus long phase. These are shifts of photometric centers, not direct measurements of conjunction-time changes.}
\label{fig:method}
\end{figure*}

$Q$ follows the local minimum, whereas $T$ asks how far a fixed, entire reference shape must be translated. Their agreement is evidence that the photometric displacement is not purely an artifact of choosing the local minimum; it does {not} establish a change of conjunction time. Conversely, disagreement can reveal a changing profile. Figure~\ref{fig:method} follows two long-phase profiles of OGLE-LMC-ECL-05935 from the original $I$ measurements through both fitted centers to their long-phase curves. In its upper panels, $N_{\rm phot}$ counts measurements within $\pm0.12$ orbital cycles and $N_{\rm conj,Q}$ counts calculated conjunctions supplying at least one point to the {local Q window}; neither number counts complete observed eclipses. The plotted points and fits are in magnitudes, while the lower panel reports center displacements in orbital cycles; it is not a plot of $(\OC)/\sigma$.

\subsection{Uncertainty, long-cycle tests, and profile shape}

The uncertainty of a composite profile must account for measurements acquired during the same orbital cycle. I therefore resampled {complete calculated conjunctions}, retaining all their points together, with 250 bootstrap realizations; at least 50 valid realizations were required for a center error. A valid $Q$ realization must again have bilateral coverage, positive curvature, and an interior minimum, while $T$ must converge within its permitted shift range. Conjunction-level grouping defines the center errors reported here; it is part of the measurement procedure, since some individual errors and selections depend on the resampling unit. The baseline $T$ template stays fixed in this resampling. Its quoted random error therefore does not, by itself, include uncertainty in constructing the reference profile $P_{\rm ref}$; I address that dependence explicitly below rather than adding an uncalibrated error term in quadrature.

For each eclipse and estimator, I fitted the long-phase dependence of valid centers with
\begin{equation}
 y(\phL)=c+a\cos(2\pi\phL)+b\sin(2\pi\phL),\quad
 K=(a^2+b^2)^{1/2}.
 \label{eq:harmonic}
\end{equation}
Here $y$ is $x_Q$ or $\delta_T$, and $(a,b)$ describes its first-harmonic modulation; $K$ is a semiamplitude, not a disk parameter. I distinguish a curve for which the harmonic can be fitted from one meeting the additional coverage and robustness requirements for detection, and from one finally selected. Table~\ref{tab:referee_sensitivity} reports these stages separately. Omitting each long-phase interval in turn checks whether one interval dominates the amplitude; no individual refit had to pass significance selection. The coefficient-pair test uses 2000 bootstrap realizations, distinct from the 250 used for each center error. I corrected its probabilities with the Benjamini--Hochberg procedure, separately for $Q$ and $T$ at a false-discovery rate (FDR) of 0.05 \citep{Benjamini1995}.\footnote{{A timing curve is selected only if it meets the coverage and stability requirements and its Benjamini--Hochberg-adjusted $p$ value satisfies $q_{\rm FDR}\leq0.05$. The correction is applied separately to the $Q$ and $T$ test families. Under the procedure's assumptions, this controls the expected fraction of false selections among selected curves within each family, not among the combined list of systems. A system is counted
when at least one of its curves is selected. A lower threshold would make selection more conservative, potentially excluding weak real modulations; a higher threshold would allow more selections while tolerating a larger false-discovery fraction.}} 
This controls the expected fraction of false selections within each family of statistical tests, under its assumptions; it is not a probability of orbital-period change or a guaranteed 5\% false-system fraction in their union. I write $q_{\rm FDR}$ for the adjusted value and reserve $q=M_{\rm d}/M_{\rm g}$ for mass ratio. These tests condition on the adopted periods and phase origins.

It is useful to compare the two eclipses without treating either as a dynamical clock. With $(a_0,b_0)$ and $(a_{0.5},b_{0.5})$ from Eq.~\eqref{eq:harmonic}, I describe the mean and half-difference of their harmonic responses as
\begin{equation}
 \boldsymbol{k}_{\rm mean}=\frac{\boldsymbol{k}_0+\boldsymbol{k}_{0.5}}{2},\qquad
 \boldsymbol{k}_{\rm diff}=\frac{\boldsymbol{k}_{0.5}-\boldsymbol{k}_0}{2},
 \quad \boldsymbol{k}_j=(a_j,b_j).
 \label{eq:paired}
\end{equation}
These mean and difference coefficients describe the photometric center curves; a significant mean can arise mostly from one eclipse, and an undetected difference does not establish equality. The uncertainties of the two eclipse curves may be correlated because they use photometry from the same system and share period and long-phase estimates. The cross-eclipse covariance of their fitted coefficients was not measured. The archived significance estimates for the mean and difference in Eq.~\eqref{eq:paired} set that covariance to zero, so their reported probabilities cannot establish that the two eclipse responses are equal or that a shared term is dynamical. I therefore do not interpret the mean component as an orbital clock.

To describe changes beyond a local center measurement, I fitted four fixed functions derived from the reference: its eclipse shape $B$, phase derivative $D$, a residual wing term $O$, and a width term $W$ \citep{RamsaySilverman2005}. The derivative resembles a translation because $P_{\rm ref}(x-\delta)\simeq P_{\rm ref}(x)-\delta P'_{\rm ref}(x)$, but its coefficient is not $T$. Appendix~\ref{app:sensitivity} specifies how overlap with the reference and derivative is removed from the $O$/$W$ trial functions. These residual terms describe additional profile structure to first order, not separate physical sources of light. Gaussian weights of width 0.075 long cycles combine nearby phases; adjacent profiles share observations, whose correlated errors are retained in joint resampling. I record a broad reference-profile indicator, selected when $D$, $O$, $W$, or the wing contrast below varies, and report $O$/$W$ separately. The broad indicator includes displacement as well as deformation. None of these tests is statistically independent of the timing measurements or defines a physical class.

To assess whether the two observed eclipse wings are balanced about the fixed, calculated center, I compared magnitudes at equal positive and negative phase offsets, $m(+u)$ and $m(-u)$, wherever both sides were sampled. A symmetric eclipse centered on the ephemeris would give zero difference. I formed the signed mean half-difference $A_{\rm mirror}$ (mag), weighting each offset by the absolute eclipse signal in the fixed reference profile so that the deeper part of the eclipse contributes more. I divided $|A_{\rm mirror}|$ by the reference eclipse depth, also in magnitudes, to compare systems with different eclipse depths; the ratio is not a fractional flux contribution. This diagnostic describes the net wing imbalance about the {calculated} center and helps assess when the local $Q$ minimum may be sensitive to the fitting window. It cannot by itself distinguish a shifted but otherwise symmetric eclipse from a changing profile, and differences of opposite sign at different offsets can cancel. Because adjacent weighted long-phase profiles share measurements, I first summarized each object and eclipse and then the sample, rather than treating every profile as independent.

\subsection{Sensitivity to the $Q$ window and the $T$ reference}\label{sec:sensitivity}

The parabola is a practical local approximation, but a disk eclipse may have an asymmetric minimum. I therefore retained the predeclared 0.75, 1.00, and 1.25 $Q$ windows and, as a directed additional test, fitted a cubic in relative flux. The cubic supplies the next local asymmetric term; it is not presumed to be a better physical eclipse model. It was accepted only when it gave a unique, interior minimum with positive curvature. Differences in both center and evaluability were retained, rather than selecting whichever procedure maximized the number of detections.

For $T$, omitting the measured long-phase interval from its reference profile tests self-contribution, but it creates a {different} reference for each interval. I therefore made a second, fixed-reference test. Conjunctions were divided deterministically into two halves, stratified by long phase and OGLE era where possible. A template constructed from half A was held fixed while all intervals in half B were measured; the roles were then reversed. To isolate the effect of the {template alone}, I also measured the same target-half points once against this independent template and once against the global template. The paired difference is a diagnostic of reference choice; a split-half series with insufficient photometry is marked unevaluable. The number of selected curves in this half-sample check is not a new detection fraction: the reduced photometry makes many otherwise testable $T$ curves unevaluable.

All alternative errors were recomputed with conjunction-level resampling and the same long-phase and FDR rules as the baseline. For the split templates, each bootstrap realization rebuilt the independent reference; this captures a source of uncertainty absent from the baseline fixed-template error. The sensitivity table in Appendix~\ref{app:sensitivity} reports both selected curves and those still testable. Its scenarios are method checks, not competing estimates of an intrinsic DPV incidence.

\subsection{Matched eclipse colors}\label{sec:color_method}

The sparse $V$ observations can nevertheless measure how the unresolved color changes during eclipse. I used existing pairs in which a $V$ epoch is bracketed by two $I$ observations from the same OGLE campaign, each within 0.5 d, and formed $C_{\rm obs}=V-I_{\rm int}$ by linear interpolation. An eclipse pair lies within 0.01 orbital cycles of its calculated center. To limit interpolation across a curved eclipse, the preferred subset additionally requires each $I$ endpoint within 0.01 orbital cycles of $V$. I compare this subset with the original brackets rather than assuming their interpolation error negligible.

Control pairs must have all three epochs outside conservative regions around both conjunctions. Their half-width is the larger of twice the nominal Q half-width and the archived full width at half eclipse depth; these masks do not measure eclipse contacts. A second check uses only epochs within 0.05 cycles of the same quadrature, 0.25 or 0.75. A control median requires at least three pairs in the same 0.1-wide long-phase interval as an eclipse pair; missing controls are not extrapolated. I also estimate how the global $I$ reference's curvature changes a linear interpolation at the three epochs. This is a sensitivity diagnostic, not an exact correction to an instantaneous eclipse.

I present the colors at two complementary levels. The first is the measured color and the map-corrected color
\begin{equation}
 C_{\rm dered}^{\rm map}=C_{\rm obs}-E(V-I)_{\rm map}(\alpha,\delta),
 \label{eq:color}
\end{equation}
using the OGLE red-clump map of \citet{Skowron2021}. Its asymmetric near/far-side dispersions are shared sightline systematics, not independent epoch errors; the corrected color is not the intrinsic color of an isolated star. The complementary level compares eclipse and control colors, or the two eclipses directly, $\Delta C_{0-0.5}=C_0-C_{0.5}$, where these subscripts denote orbital phase. Constant sightline reddening cancels, but different sampling of long phase does not. Summary errors use 5000 fixed-seed resamples of groups connected by a shared $I$ endpoint, with common noise draws for reused $I$ measurements and propagated $V/I$ errors. They remain conditional on the adopted clocks, sampled epochs and linear interpolation; reference-curvature differences are assessed separately. Color alone does not identify the occulting component.

\section{Observed long-cycle variations}\label{sec:observed}

\subsection{Eclipse-center modulations and the two eclipses}\label{sec:timing_results}

The 55 curated systems yield 1196 valid $Q$ and 1218 valid $T$ centers in non-overlapping long-phase intervals. These counts refer to composite profiles, not individual orbital eclipses. Of the 55, 49 have sufficient data for at least one harmonic test. Twenty-seven of these 49 show at least one $Q$ or $T$ modulation selected after the coverage, robustness, and FDR criteria, comprising 46 selected curves. The fraction $27/49=0.551$ characterizes this curated, evaluable experiment; its 95\% Wilson interval is 0.413--0.681. Because the input photometry differs strongly in cadence and quality among objects, it does not measure the fraction of all Magellanic DPVs with changes in orbital period or conjunction time.

\begin{figure}
\centering
\includegraphics[width=\columnwidth]{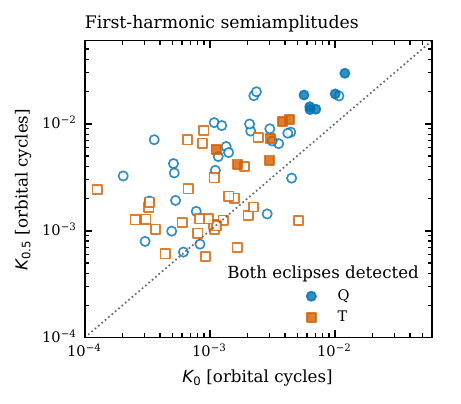}
\caption{{First-harmonic semiamplitudes of the observed photometric eclipse-center variations at orbital phases 0 and 0.5, expressed in orbital cycles. Each point represents one system measured with the same estimator in both eclipses: circles denote \(Q\), and squares denote \(T\). Filled symbols identify systems with a modulation selected in both eclipses using that estimator; open symbols do not meet this two-eclipse selection condition. The dotted line marks equal semiamplitudes. The comparison describes photometric-center variations, not directly measured changes in stellar conjunction times.}}
\label{fig:population}
\end{figure}

At phase 0, $Q$ and $T$ each select seven of 32 testable curves. At phase 0.5, $Q$ selects 22 of 47 and $T$ selects ten of 49. Fifteen systems at phase 0.5 are selected by $Q$ but not $T$, against three selected by $T$ but not $Q$ (exact paired-test $p=0.0075$, $q_{\rm FDR}=0.030$ across four such comparisons). Here $p$ is the unadjusted probability, under equal chances of $Q$-only and $T$-only selection among discordant systems, of an imbalance at least this large; $q_{\rm FDR}$ is the value adjusted for the four comparisons. This difference is expected if a moving local minimum and a translated full profile respond differently to shape changes. It also warns against calling every selected photometric shift an orbital-clock change. Among selected curves only, the median $Q$ semiamplitudes are 0.00633 and 0.01193 orbital cycles at phases 0 and 0.5, corresponding to 59.9 and 127.6 min after each system's period is applied. The corresponding $T$ medians are 0.00300 and 0.00719 cycles, or 27.2 and 69.4 min. These are descriptive scales of detected curves, not an unbiased amplitude distribution. {To compare the strength of the photometric timing modulation in the two eclipses, I plot their first-harmonic semiamplitudes against each other in  Fig.~\ref{fig:population}. Each point pairs the phase-0 and phase-0.5 measurements of one system obtained with the same estimator. Many points lie above the equality line, indicating larger semiamplitudes at phase 0.5 among the systems shown. This amplitude comparison does not establish whether the two eclipse-center curves vary in phase or whether the stellar conjunction times change.}

OGLE-LMC-ECL-05935, 12848, and 32315 each have individually selected $Q$ and $T$ curves at both eclipses. Their responses are not identical: a differential $Q$ component is selected for each under the working approximation of zero cross-eclipse covariance. The data do not provide a directly measured cross-eclipse covariance or a predeclared physical tolerance within which the two curves could be considered equivalent. These objects merit follow-up as two-eclipse timing cases, but neither prove nor exclude a change in actual conjunction time. In OGLE-LMC-ECL-17632, by contrast, the strong $Q$ modulation is confined to phase 0.5; a mean of the eclipse coefficients would obscure that distinction (Fig.~\ref{fig:observed_examples}). Further observed curves are presented in Appendix~\ref{app:examples}.

\begin{figure*}
\sidecaption
\includegraphics[width=12cm]{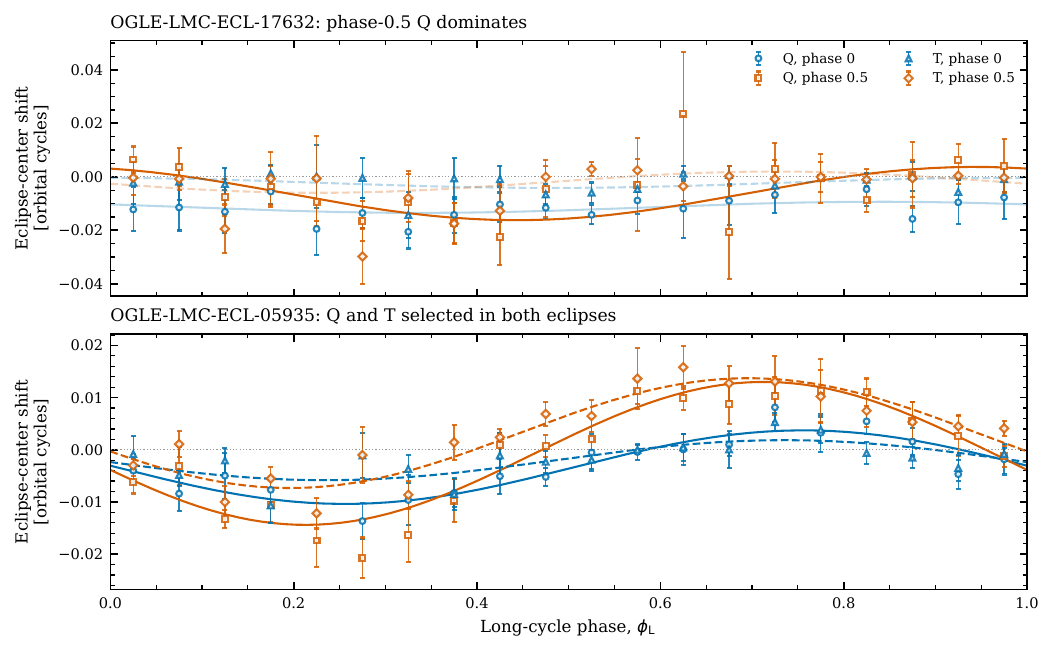}
\caption{Two observed patterns of long-cycle photometric eclipse-center variation. $Q$ and $T$ measurements are shown separately for the operational eclipses at phases 0 and 0.5, with conjunction-level bootstrap errors and first-harmonic fits; faint curves fail the individual selection criteria. OGLE-LMC-ECL-17632 is dominated by $Q$ at phase 0.5. OGLE-LMC-ECL-05935 has selected $Q$ and $T$ curves at both eclipses, but they differ and do not establish a shared dynamical clock.}
\label{fig:observed_examples}
\end{figure*}

For the selected phase-0.5 curves, the rising crossing of the fitted $Q$ harmonic through its mean is concentrated near long phase 0.45--0.50, with circular mean 0.490 for 22 curves. The ten selected $T$ curves have a broader distribution and mean 0.430. Rayleigh probabilities for uniform rising phases are $1.7\times10^{-6}$ and $4.3\times10^{-3}$, respectively. Each falling crossing follows from the same harmonic half a cycle later and is not an independent test. These distributions condition on period reliability, coverage, detectable amplitude and FDR selection. Without injection and recovery through the OGLE sampling, I cannot infer the crossing distribution of all Magellanic DPVs.

\subsection{Profile changes and sensitivity of the centers}

An eclipse can change shape without a measurable center modulation, while a translated symmetric eclipse changes its profile relative to the fixed calculated center. I therefore distinguish timing from a broader reference-profile indicator. Among the 49 systems admissible for a timing test, 20 have both indicators selected, seven only timing, six only the broad profile indicator, and 16 neither. Across all 55, the broad indicator is selected in 30, including four without sufficient timing data. This is not a count of 30 independently established deformations: its derivative and mirror terms can respond to translation alone. Separately, ten systems have a selected $O$ coefficient, eight also with selected timing; $W$ is selected in none. These residual wing terms provide more specific evidence of additional profile structure within the first-order representation, but not a unique physical interpretation. Table~\ref{tab:referee_selected55} distinguishes the broad and $O$/$W$ indicators.

The observed eclipse profiles are often unequal on opposite sides of the fixed, calculated center. Across the 55 systems, the median per-system mirror-asymmetry-to-depth ratio is 0.0298 at phase 0 and 0.0619 at phase 0.5; the respective 90th percentiles are 0.0978 and 0.1936. These ratios compare differences in magnitudes with eclipse depth; they are not fractions of disk light or a separation of genuine center shifts from profile changes. The 1056 measured profiles overlap in long phase and are not independent trials. Such mirror differences can accompany a shifted minimum and make a local parabolic fit sensitive to its window even when its formal fit is precise.

The limited procedural changes quantify this sensitivity (Table~\ref{tab:referee_sensitivity}). A cubic gives a valid $Q$ center for 1151 of 1196 baseline profiles; where both fits work, the median absolute difference is 0.00337 cycles and the 90th percentile 0.01195. Omitting the measured interval from the $T$ reference leaves 1215 of 1218 profiles valid, with median difference 0.000772 cycles and 90th percentile 0.00588. The cubic scenario selects 19 systems versus the baseline 27, losing nine and gaining one. All nine lost systems still have an adjustable formerly selected harmonic; seven remain admissible but no longer pass FDR, while two fail admissibility. Thus the losses are not merely missing-data effects. The $T$ omission scenario selects 26. These are measurement checks, not alternative population rates.

Nevertheless, twelve $Q$ curves in ten systems retain selection with the baseline parabola, the cubic and all three tested windows: two phase-0 curves and ten phase-0.5 curves. The systems are OGLE-LMC-ECL-05935, 06957, 11994, 12848, 14202, 17632, 23767, 30826, 32315 and OGLE-SMC-ECL-01807; only 05935 and 32315 retain a phase-0 selection throughout. This persistence supports the existence of the phenomenon beyond a particular local function. It is a descriptive intersection of controls sharing data, not a newly calibrated detection fraction.

The independent-template test makes the $T$ dependence easier to isolate: it fits the same target points using two different reference profiles. Among 2390 comparable measurements, the median absolute change ranges from 0.00116 to 0.00255 orbital cycles across the two eclipses and sample halves; the 90th percentiles range from 0.00684 to 0.01256 cycles. The signed median is near zero, but individual centers can shift substantially. In 227 comparisons, the observed difference lies more than one bootstrap standard deviation from the bootstrap median, so that spread is not a dependable formal confidence interval in every case. Using only half the observations also leaves just 17 or 20 of the 90 baseline-testable $T$ curves measurable. The number selected in this test is therefore not comparable with the 27/49 baseline yield. Template choice matters particularly when interpreting a small $Q$--$T$ difference in an individual system; it cannot be represented by one extra uncertainty common to all systems.

Among the detected curves, I find no relation of timing semiamplitude with $\Porb$, $\Plong$, or their ratio that remains significant after allowance for the multiple comparisons. This does not show that the underlying physical amplitude is period-independent: only systems with adequate sampling and detectable shifts enter this comparison.

Table~\ref{tab:template_objects} provides the reference sensitivity separately for every system and eclipse. For OGLE-LMC-ECL-05935, the median absolute paired $T$ differences are 0.00217 and 0.00420 cycles at phases 0 and 0.5, with 90th percentiles 0.00561 and 0.01157. At phase 0.5, the corresponding medians are 0.00351 for 12848 and 0.00246 for 32315. These object-dependent scales must be considered before interpreting a small $Q$--$T$ difference. They are not systematic standard errors; the baseline $T$ selection still conditions on its fixed reference, and the full reference uncertainty has not been propagated into that selection.

\subsection{Eclipse colors}\label{sec:colors_result}

The existing catalog has 2760 matched $V-I$ pairs, including 66 near phase 0 and 47 near phase 0.5. The local-interpolation requirement retains 51 and 41, respectively, giving an eclipse color for 27 objects and both colors for 16, all in the LMC (Table~\ref{tab:colour_summary}). Only two have at least three local pairs in each eclipse. The other pairs are not automatically outside eclipse: the conservative mask admits 1161 controls and the quadrature-only check 538. Missing regions remain unmeasured.

\begin{table*}\caption{Coverage and sensitivity of the eclipse-color analysis.}\label{tab:colour_summary}\centering\small\begin{tabular}{@{}lrl@{}}\hline\hline Quantity & Result & Scope \\\hline
Existing matched pairs & 2760 & same photometric inputs \\
Local eclipse pairs: phase 0 / 0.5 & 51 / 41 & 92 retained of 113 \\
Objects with an eclipse color & 27/55 & missing colors not imputed \\
Both eclipses, local brackets & 16 & two have three pairs in each \\
Phase 0 redder, local brackets & 14/16 & signs, not individual detections \\
Median $\Delta C_{0-0.5}$, local & 0.153 mag & sightline reddening cancels \\
Both eclipses, original brackets & 18 & 16 redder; median 0.125 mag \\
Conservative / quadrature controls & 1161 / 538 & phase masks include I endpoints \\
\hline\end{tabular}\tablefoot{The preferred local subset uses the same $V$/$I$ measurements as the original analysis, with a tighter phase separation for interpolation. Control masks do not measure eclipse contacts. Both inter-eclipse summaries remain conditioned on their sampled long phases.}\end{table*}

\begin{figure*}
\centering
\includegraphics[width=0.88\textwidth]{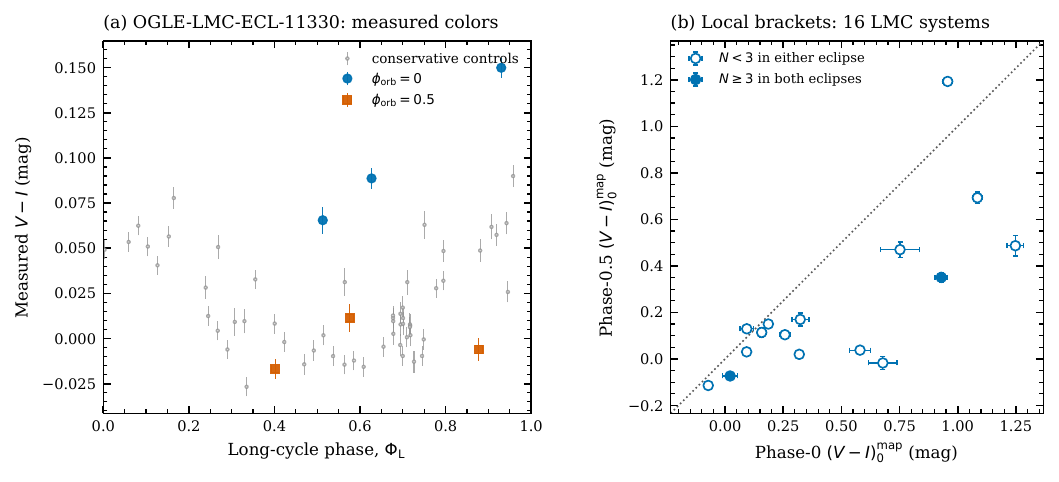}
\caption{Colors from locally bracketed eclipse pairs. Panel (a): measured $V-I$ for OGLE-LMC-ECL-11330, with conservative phase-selected controls. Subtracting the common $E(V-I)=0.067$ mag gives map-corrected colors without changing their relative pattern. Panel (b): corrected medians for the 16 LMC systems with both local eclipse colors; the diagonal is equality and filled symbols mark at least three pairs in both eclipses.  {Here \(I\) denotes the OGLE \(I\) passband. The error bars include photometric measurement errors and account for the reuse of the same \(I\)-band measurements in different color estimates. They exclude the common reddening-map uncertainty for each system and additional interpolation errors arising from changes in the instantaneous eclipse profile. The common reddening correction cancels from the color difference between the two eclipses.} The two eclipses need not sample identical long phases.}
\label{fig:observed_colours}
\end{figure*}

Fourteen of the 16 locally matched systems have a redder phase-0 median, with median inter-eclipse contrast 0.153 mag. The original brackets give 16 of 18 and 0.125 mag. These signs are not 14 or 16 individually significant detections. Figure~\ref{fig:observed_colours} and Table~\ref{tab:app_colours} present the preferred local colors and uncertainties accounting for reused I points. For these pairs, the global-reference estimate of absolute interpolation bias has median 0.00092 mag and 90th percentile 0.00983 mag, versus 0.00184 and 0.02260 mag for the excluded eclipse brackets. The reference is not the instantaneous eclipse, so this check reduces rather than eliminates the interpolation concern.

For OGLE-LMC-ECL-11330, the locally bracketed phase-0 pairs give a median excess of 0.081 mag against conservative controls and 0.074 mag against quadrature controls in matched long-phase intervals; $E(V-I)=0.067$ mag shifts every color together. The tendency toward a redder deeper eclipse survives these limited checks and is compatible with a cool donor in front preferentially suppressing blue gainer/inner-disk light. It does not identify that orientation: a cool rim, stream material, or changing circumstellar emission can give similar unresolved colors. Sightline reddening cancels from the inter-eclipse difference, but sparse V sampling, different sampled long phases and the all-LMC subset remain important limits.

\section{An illustrative DPV with fixed conjunctions}\label{sec:simulation}

\subsection{Numerical experiment and phase convention}

The observed eclipse-center changes invite a specific countercheck: can a variable distribution of circumstellar light move the measured minimum if the stars remain on the same orbit? I addressed this with {one} artificial circular, semidetached DPV, not with a fit to any of the 55 objects. The Roche-lobe-filling donor, gainer, and optically thick disk were represented by the light-curve calculation of the Djura\v{s}evi\'c modeling family \citep{Djurasevic1992a,Djurasevic1996}. It includes projected and occulted stellar and disk surfaces, limb and gravity darkening, irradiation, disk self-occultation, and optional localized hot and bright rim regions.

{The parameters in Table~\ref{tab:synthetic}  specify an illustrative numerical configuration, not an observationally fitted binary. Fixed entries are prescribed inputs, whereas the donor mass, orbital separation, and outer disk temperatures follow from the adopted mass ratio, Kepler’s law, and disk temperature law, respectively. The tabulated digits specify or summarize these numerical values; they do not represent measurement precision or formal uncertainties. Rounding the displayed values does not change the inputs used in the calculations. No observational parameter uncertainties were assigned or propagated, so the results remain conditional on the adopted configuration. The experiment demonstrates that photometric timing shifts can arise at fixed conjunctions; it does not establish precise predictions for an observed system or robustness against untested parameter variations.}

\begin{table}[tbp]
\caption{Adopted parameters of the single artificial DPV.}
\label{tab:synthetic}
\centering\small\setlength{\tabcolsep}{3pt}
\begin{tabular}{@{}>{\raggedright\arraybackslash}p{0.31\columnwidth}>{\raggedright\arraybackslash}p{0.35\columnwidth}>{\raggedright\arraybackslash}p{0.26\columnwidth}@{}}
\hline\hline
Quantity & Value & Role \\
\hline
$\Porb$ & 7.0 d & fixed \\
$M_{\rm g},M_{\rm d}$ & $7.0,1.61\,M_\odot$ & fixed, derived \\
$q=M_{\rm d}/M_{\rm g}$ & 0.23 & fixed \\
$a$; $i$ & \(a\simeq31.6\,R_\odot\); $82^\circ$ & derived; fixed \\
$T_{\rm g},T_{\rm d}$ & 20\,000, 7000 K & fixed \\
$R_{\rm in}$ & $3.024\,R_\odot$ & fixed \\
$T(r)$ & $20\,000(r/R_{\rm in})^{-1/2}$ K & disk law \\
$T_{\rm out} (K)$ & \(1.25\times10^4,\ 8.86\times10^3\)& two disk radii \\
$R_{\rm d}/\Rt$ & 0.5, 1.0 & varied \\
$\Rt/a$ & $0.6/(1+q)=0.488$ & adopted scale \\
$\beta$ & $0^\circ,1^\circ,3^\circ,7^\circ$ & varied \\
$H_{\rm in},H_{\rm out}$ & $1.0,2.5\,R_\odot$ & total thickness \\
Hot/bright rim factors & 1.20, 1.10 & on/off together \\
Rim-region widths & $20^\circ,30^\circ$ & full azimuthal widths \\
\hline
\end{tabular}
\tablefoot{The outer temperatures correspond respectively to $R_{\rm d}/\Rt=0.5$ and 1.0. $\Rt$ is the adopted nominal tidal radius. The rim factors multiply the local temperature, not the integrated disk flux. The hot region is anchored to the first ballistic-stream intersection with the outer rim; the bright region lies opposite it. Their tabulated widths are full widths, not angular radii. Both regions are switched on or off {together}; their separate effects were not computed. In this artificial system the donor is in front at orbital phase 0; this component assignment is not imposed on the observed sample. The coplanar and inclined families also differ in their prescribed thickness evolution.}
\end{table}

The coplanar comparator has a prescribed thickness cycle, $H_j(\Phi)=H_{j,0}[1+0.40\cos(2\pi\Phi)]$, with $\Plong=33\Porb$. For an inclined disk I held thickness fixed and rotated its plane rigidly about the orbital axis. The illustrative precession period followed the ring approximation used by \citet{Jiao2026},
\begin{equation}
 \frac{P_{\rm prec}}{\Porb}=\frac{4}{3}\frac{\sqrt{1+q}}{q}
 \left(\frac{a}{R_{\rm d}}\right)^{3/2}\frac{1}{\cos\beta}.
 \label{eq:precession}
\end{equation}
It is an assumed modulation rate, not a period inferred from OGLE. The two radii correspond to about 53.4 and 18.9 orbital periods. Geometric phase $\Phi=0$ denotes maximum projected area of the axisymmetric tilted disk. With $+x$ from gainer to donor, its normal is $(\sin\beta,0,\cos\beta)$ there; increasing $\Phi$ follows the adopted retrograde sense. To compare phase relations with observations, I independently set $\phi_{\rm L}=0$ for each model sequence at its maximum computed $I$-band flux at orbital phase 0.25. No further free phase shift was fitted. The model quadrature maximum need not be identical to the observed maximum of a full-orbit long-cycle light curve.

For each radius, tilt, and joint rim-region state, I calculated twelve long phases. The 16 resulting model sequences contain 192 physical configurations and 384 eclipse profiles. I measured $Q$ and $T$ from each noise-free profile in a $\pm0.08$ orbital-cycle window about the {fixed} conjunction, at a 0.002-cycle grid spacing, using projected-area occultations of discretized Roche and disk surfaces. The stellar conjunction phases remained 0 and 0.5 throughout. I also measured eclipse depth, width, asymmetry, and the model $I$ flux at orbital phase 0.25. Repeating the timing extraction at 0.004-cycle spacing changed $Q$ and $T$ by no more than 1.87 and 1.54 min. This checks the adopted timing grid, not convergence of every stellar-surface and disk element.

\subsection{What the artificial system produces}

\begin{figure*}
\sidecaption
\includegraphics[width=12cm]{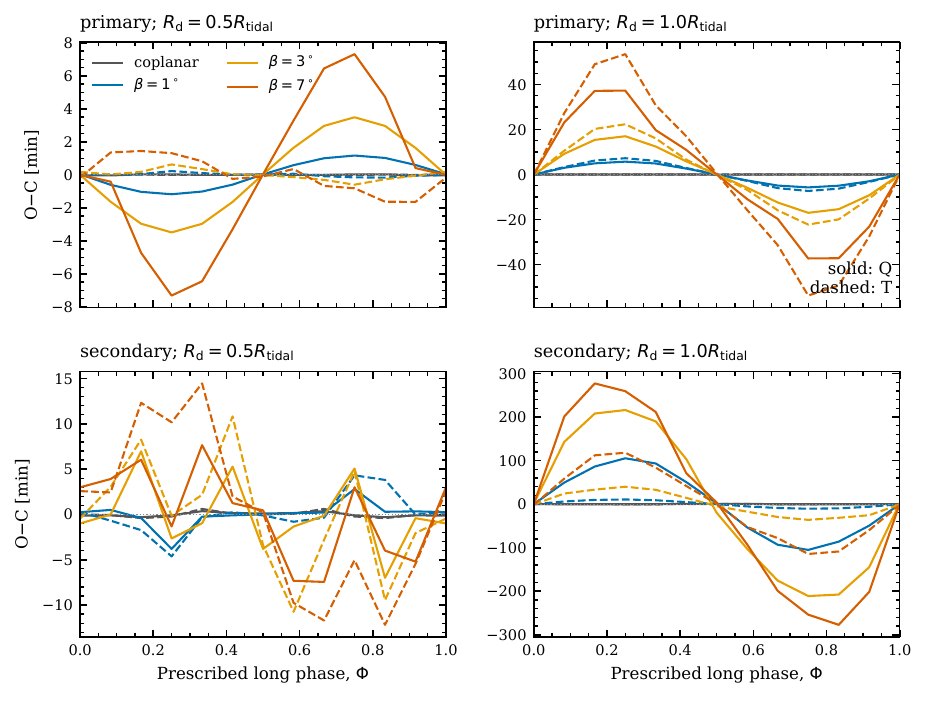}
\caption{Noise-free, mean-centered photometric eclipse-center curves of the spot-free artificial DPV. Gray curves represent the coplanar model with a prescribed thickness cycle; colored curves represent rigidly inclined disks at fixed thickness. Conjunction remains at orbital phases 0 and 0.5. The distinction between these families is not a tilt-only experiment because their thickness laws differ.}
\label{fig:sim_oc}
\end{figure*}

Figure~\ref{fig:sim_oc} shows $Q$ and $T$ center changes when the localized rim regions are off; the disk and its baseline rim emission remain present. The coplanar thickness cycle leaves the phase-0 $Q$ minimum almost stationary (peak-to-peak below 0.07 min) and moves the phase-0.5 minimum by 0.67--1.12 min. With $R_{\rm d}=0.5\Rt$, a $7^\circ$ tilt yields $Q$ peak-to-peak changes of about 15 min. At $R_{\rm d}=\Rt$, the phase-0.5 $Q$ excursions reach 210, 427, and 554 min for tilts of $1^\circ$, $3^\circ$, and $7^\circ$. The corresponding $T$ peak-to-peak changes reach 21, 76, and 232 min. Switching both rim regions on changes some coplanar sequences markedly, including a 303-min phase-0.5 $Q$ excursion for the large disk; because the regions were toggled together, neither one's contribution is isolated. Large shifts with both regions off show that these added regions are not necessary for the effect in this particular model.

Even without observational noise, $Q$ and $T$ can disagree. Their maximum instantaneous separation is 176 min in a large-disk phase-0.5 eclipse, where the profile is strongly asymmetric. Figure~\ref{fig:sim_qt} in Appendix~\ref{app:model} shows the point-by-point $Q$--$T$ comparison. The calculation is therefore a concrete counterexample to equating any photometric O$-$C value with a moving stellar conjunction. It does not imply that the same geometry produced an observed OGLE shift.

\begin{figure*}
\sidecaption
\includegraphics[width=12cm]{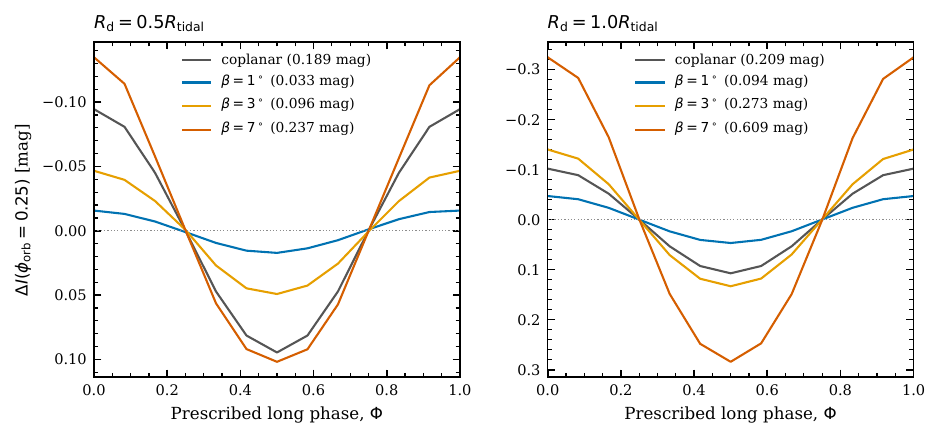}
\caption{Modeled $I$-band brightness at orbital phase 0.25 for the spot-free artificial system. Each sequence is centered on its own median over twelve long phases; legends give its peak-to-peak range. Gray denotes coplanar variable-thickness cases and colors the fixed-thickness inclined disks. The fluxes share the forward geometry used for Fig.~\ref{fig:sim_oc}, but have not been fitted to OGLE photometry.}
\label{fig:sim_quad}
\end{figure*}

The same artificial configurations change brightness outside eclipse (Fig.~\ref{fig:sim_quad}). With the rim regions off, the peak-to-peak model variation at phase 0.25 is 0.189--0.209 mag in the two coplanar radius families, 0.033--0.237 mag for inclined disks of radius $0.5\Rt$, and 0.094--0.609 mag for inclined disks of radius $\Rt$. This is a single-passband quadrature prediction, not an observed OGLE range or a modeled $V-I$ color.

The phase of a timing modulation supplies a further comparison beyond its amplitude. After applying the stated brightness-maximum convention, the first-harmonic fits to large tilted-disk sequences with $\beta=3^\circ$ and $7^\circ$ place the rising phase-0.5 $Q$ and $T$ mean crossings close to 0.0 and the falling crossings near 0.5. The selected observed $Q$ harmonics instead rise near 0.45--0.50, while $T$ has a broader distribution (Sect.~\ref{sec:timing_results}). Small-disk and coplanar {full modulation curves} have a broader range of crossings and can cross their mean more than twice. That behavior is not equivalent to the single rising crossing of a first harmonic. Thus the large-disk examples produce conspicuous eclipse-center shifts but do not reproduce the observed phase relation under the adopted convention.

\section{Discussion}\label{sec:discussion}

\subsection{A measured photometric phenomenon, not yet a measured mechanism}

The central observational result is the detection of long-cycle-dependent eclipse-center variations in 27 of 49 admissible, curated Magellanic DPVs under the adopted procedure. Ten systems retain a selected $Q$ curve across all tested local functions and windows, showing that the existence of the phenomenon does not rest on a parabola alone. These measurements reveal variations of photometric centers, not by definition stellar conjunction times. The two eclipses, both center definitions and the complementary profile diagnostics describe a property that explanations of the long cycle must address in the affected systems.

The broad profile indicator accompanies timing selection in 20 admissible systems, but it includes responses to translation itself. More specifically, eight timing-selected systems also have a selected residual wing coefficient $O$, consistent with additional profile structure within the adopted linear representation. This does not identify the changing structure. $Q$ and $T$ can disagree, and template choice introduces further method dependence. Conversely, detections by both methods at both eclipses, as in three objects, deserve attention without proving a common conjunction shift. Their cross-eclipse covariance is insufficiently measured for an equivalence claim. Genuine conjunction-time variation is neither demonstrated nor ruled out in an individual DPV; the circular reference and long-cycle harmonic do not distinguish secular evolution, light-travel time or apsidal effects.

The color result gives a separate clue: phase 0 is redder in 14 of 16 locally matched systems, compared with 16 of 18 under the original interpolation criterion. Preferential loss of blue gainer/inner-disk light when the donor is in front is one explanation, but cooler circumstellar material can also contribute. The limited checks support the descriptive color tendency, not a unique orientation or a population incidence. A constant sightline correction cancels, whereas interpolation and long-phase sampling remain potential systematics. The single-band numerical experiment does not test the $V-I$ result directly.

\subsection{What the numerical comparison does and does not show}

The artificial DPV answers a deliberately narrower question than the OGLE analysis. Its conjunctions are fixed by construction, yet its changing disk projection and emission generate large photometric O$-$C shifts, eclipse-shape changes, and variations in quadrature brightness. It demonstrates physical {possibility}: shifts of the measured minimum need not be shifts of the stellar orbit. It neither fits an observed binary nor measures the frequency of disk tilt, precession, or rim variability among DPVs in the Magellanic Clouds.

For a meaningful comparison, I distinguish the same measure of timing variation in observations and models. 
{Across the calculated configurations, the largest first-harmonic semiamplitudes reach approximately 35 and 270 min for \(Q\), and 48 and 110 min for \(T\), at orbital phases 0 and 0.5, respectively. Other configurations yield much smaller fitted amplitudes. The smallest values have not been established as numerically resolved nonzero signals and are not used in the physical interpretation. These synthetic amplitudes overlap} part of the scale seen in the observed curves. Nevertheless, the observed median describes systems that passed detectability tests, whereas the 16 variants of one artificial system were not subjected to OGLE cadence, photometric errors, period ambiguity, or the observational selection procedure. A model's full peak-to-peak excursion also cannot be equated with a first-harmonic semiamplitude. The overlap therefore supports the relevance of the counterexample, not an estimate of a disk parameter.

The phase relation is a useful limitation of this particular example. The large inclined disks produce secondary-eclipse mean crossings about half a cycle from the concentration of the {selected} observed curves, under the explicitly stated brightness-maximum convention. These realizations therefore do not reproduce the observed phase relation even though they can produce large timing shifts. This does not reject inclined disks as a class: radius, rim emission, viewing geometry, and the mapping between quadrature and observed long-cycle maxima can matter, while selection may itself sharpen the observed crossing distribution. The coplanar comparator also has a variable thickness whereas the inclined disks keep thickness fixed. Their difference cannot be attributed to inclination alone. Similarly, the two rim regions were switched together, so the experiment does not isolate either one.

OGLE-LMC-DPV-062 (OGLE-LMC-ECL-12848 here) already has a separate disk-evolution analysis \citep{Mennickent2026DPV062}; an O$-$C analysis of OGLE-BLG-ECL-275974 has been submitted. Such object-specific studies can connect center shifts to fuller orbital light curves, spectroscopy, and physical disk models. Particularly discriminating next calculations would hold thickness fixed while changing tilt, vary each rim region separately, synthesize both $V$ and $I$ throughout the orbit, and pass synthetic light curves through the actual OGLE sampling. Those are proposed tests, not results of the present experiment. For the observed objects, better sampling of both eclipses and independent constraints on the stellar orbit are needed before assigning a conjunction-time variation or a specific disk geometry.

\subsection{Selection and uncertainty limits}

The 55 objects were visually curated after an automatic survey that sometimes used incorrect periods; the final working clocks and their local jackknife errors are reported, with aliases kept as cautions. The 54 other candidates cannot be counted as timing non-detections simply because their present coverage or long clock fails the analysis. The 49-system denominator likewise depends on cadence, eclipse visibility, period quality, and the timing procedure. A survey-wide incidence would require a parent DPV definition and an injection--recovery selection function through those steps. The measured concentration of secondary-eclipse crossing phases is subject to the same selection. Finally, the baseline bootstrap conditions on a fixed $T$ reference; the split-reference experiment demonstrates why small object-level $Q$--$T$ agreements or differences require the template sensitivity to be consulted rather than interpreted from the baseline error alone.

\section{Conclusions}\label{sec:conclusions}

\begin{enumerate}
\item From 109 OGLE eclipsing-DPV candidates in the Magellanic Clouds, period review and visual curation produced 55 systems with adopted orbital and long-cycle clocks and explicit coverage cautions. Forty-nine have data adequate for at least one long-phase eclipse-center test. The remaining candidates are not timing non-detections.
\item The adopted procedure detects long-cycle-dependent photometric-center variations in 27 of 49 admissible systems. Ten retain a selected $Q$ curve across all tested local functions and windows. Three have both $Q$ and $T$ selected in both eclipses, without establishing a common change of conjunction time. The broad profile indicator accompanies timing in 20 systems but includes translation; eight also have a selected residual wing term $O$.
\item The phase-0 eclipse is redder in 14 of 16 systems with locally interpolated colors at both eclipses; the original brackets give 16 of 18. A constant sightline correction cancels, but interpolation, sparse sampling and the all-LMC subset prevent a unique identification of the occulting component or a population inference.
\item One artificial DPV with fixed conjunctions produces photometric shifts from below a minute to hundreds of minutes as its prescribed disk geometry changes. It is an illustrative explanation for how such shifts {can} arise, not a fit or proof of nodal precession in the observed systems. Its large tilted-disk cases do not reproduce the selected observed secondary-eclipse crossing phases. The origin of the observed timing variations, including the possible contribution of genuine conjunction-time changes, remains open to object-specific tests.
\end{enumerate}

\begin{acknowledgements}
I thank the anonymous referee for constructive comments that helped clarify the sample selection, timing measurements, and limits of the numerical comparison. This work uses observations from the Optical Gravitational Lensing Experiment (OGLE). I acknowledge support from ANID BASAL Centro de Astrof\'{i}sica y Tecnolog\'{i}as Afines ACE210002 (CATA). OpenAI tools assisted with code development, verification, and language editing. I reviewed the results and take responsibility for the scientific content.
\end{acknowledgements}

\begin{appendix}

\onecolumn
\section{Object-level selection and measurement inventory}\label{app:inventory}

Table~\ref{tab:referee_selected55} gives adopted phase origins and the timing inventory for 55 curated DPVs. $N_I$ counts usable $I$ photometry; $N^{Q/T}_{\rm conj}$, contributing conjunctions, not complete eclipses; $C/Q/T$, covered intervals and valid centers. Missing timing information is not a non-detection.

\begingroup\footnotesize\renewcommand{\arraystretch}{0.84}\setlength{\tabcolsep}{1.25pt}
\begin{longtable}{@{}lrrrrcrrrcccccc@{}}
\caption{Curated DPVs: adopted periods, phase origins, and usable eclipse-timing data}\label{tab:referee_selected55}\\
\hline\hline
Object & $P_{\rm orb}$ & $P_{\rm long}$ & $T_0$ & $T_{\rm L,max}$ & Br. & $N_I$ & \shortstack{$N^{Q/T}_{\rm conj}$\\$\phi=0$} & \shortstack{$N^{Q/T}_{\rm conj}$\\$\phi=0.5$} & \shortstack{$C/Q/T$\\$\phi=0$} & \shortstack{$C/Q/T$\\$\phi=0.5$} & E & Tim. & Prof. & $O$/$W$ \\
\hline
\endfirsthead
\caption[]{continued.}\\
\hline\hline
Object & $P_{\rm orb}$ & $P_{\rm long}$ & $T_0$ & $T_{\rm L,max}$ & Br. & $N_I$ & \shortstack{$N^{Q/T}_{\rm conj}$\\$\phi=0$} & \shortstack{$N^{Q/T}_{\rm conj}$\\$\phi=0.5$} & \shortstack{$C/Q/T$\\$\phi=0$} & \shortstack{$C/Q/T$\\$\phi=0.5$} & E & Tim. & Prof. & $O$/$W$ \\
\hline
\endhead
\hline
\multicolumn{15}{r}{continued on next page}\\
\endfoot
\hline\hline
\endlastfoot
LMC-00778 & $8.278183(9)$ & $253.48(17)$ & 6311.40629 & 6284.98532 & V1 & 1355 & 66/96 & 170/285 & 3/3/3 & 8/8/8 & Y & Y & Y & -- \\
LMC-04725 & $7.395574(26)$ & $247.97(14)$ & 6161.69654 & 6260.22951 & V1 & 2754 & 93/162 & 123/201 & 7/7/7 & 7/7/7 & -- & -- & Y & -- \\
LMC-05532 & $3.79569^{\mathrm{u}}$ & $121.1^{\mathrm{u}}$ & 6312.76491 & 6310.57384 & V1 & 6241 & 222/420 & 305/513 & 20/19/20 & 20/20/20 & Y & -- & Y & -- \\
LMC-05935 & $5.1798920(30)$ & $153.831(30)$ & 5590.30170 & 5587.89393 & V1 & 6539 & 247/429 & 350/525 & 19/19/19 & 20/20/20 & Y & Y & Y & Y \\
LMC-06376 & $5.77173(4)$ & $203.6(35)$ & 6286.51781 & 6208.82690 & V1 & 1346 & 35/59 & 134/224 & 2/2/2 & 7/7/7 & -- & -- & -- & -- \\
LMC-06957 & $6.296876(10)$ & $208.69(16)$ & 5593.17721 & 5679.95687 & V1 & 6555 & 201/335 & 322/495 & 17/17/17 & 20/20/20 & Y & Y & Y & Y \\
LMC-07098 & $24.5117(4)$ & $761.8(25)$ & 5603.76444 & 5892.51455 & V1 & 6566 & 147/191 & 163/208 & 9/9/9 & 9/9/9 & Y & Y & Y & Y \\
LMC-07343 & $17.02401(15)$ & $561.8(21)$ & 9984.37315 & 10022.36087 & V2 & 6399 & 191/267 & 216/297 & 10/10/10 & 10/10/10 & Y & -- & Y & -- \\
LMC-09173 & $5.845748(21)$ & $174.63(5)$ & 9966.42032 & 10008.28088 & V2 & 6630 & 222/355 & 314/494 & 16/16/16 & 19/19/19 & Y & Y & -- & -- \\
LMC-11330 & $7.284300(17)$ & $177.7(5)$ & 9957.75908 & 9919.18635 & V2 & 7156 & 283/425 & 335/504 & 19/19/19 & 20/20/20 & Y & Y & Y & -- \\
LMC-11906 & $6.96714(4)$ & $207.97(17)$ & 10002.28388 & 10039.09587 & V2 & 5785 & 158/222 & 198/260 & 10/10/10 & 10/10/10 & Y & Y & -- & -- \\
LMC-11961 & $12.74797(12)$ & $760.6(32)$ & 10021.64516 & 10325.67057 & V2 & 5204 & 99/132 & 125/174 & 8/8/8 & 9/9/9 & Y & -- & -- & -- \\
LMC-11994 & $6.517427(6)$ & $206.81(6)$ & 5452.95868 & 5528.93999 & V1 & 6758 & 264/413 & 330/512 & 19/19/19 & 20/20/20 & Y & Y & Y & -- \\
LMC-12704 & $18.1557(10)$ & $347.8(5)$ & 9982.13571 & 9970.50869 & V2 & 6403 & 124/159 & 181/245 & 7/7/7 & 10/10/10 & Y & Y & Y & -- \\
LMC-12808 & $3.131011(34)$ & $114.0(5)$ & 9998.06309 & 10051.70573 & V2 & 5778 & 179/337 & 172/419 & 10/10/10 & 10/7/10 & Y & -- & Y & -- \\
LMC-12848 & $6.904854(15)$ & $229.76(9)$ & 5455.76235 & 5370.54921 & V1 & 6768 & 204/315 & 304/468 & 16/16/16 & 19/19/19 & Y & Y & Y & Y \\
LMC-13898 & $6.00730^{\mathrm{u}}$ & $213.83(34)$ & 9970.21542 & 9896.87087 & V2 & 6470 & 246/379 & 334/506 & 18/18/18 & 20/20/20 & Y & Y & -- & -- \\
LMC-14202 & $27.73311(23)$ & $1047(7)$ & 6315.47170 & 6272.16456 & V1 & 6239 & 75/100 & 121/145 & 6/6/6 & 8/8/8 & Y & Y & Y & -- \\
LMC-14413 & $38.1590(4)$ & $1545(8)$ & 9971.68057 & 9317.93212 & V2 & 6797 & 99/125 & 137/160 & 8/8/8 & 10/10/10 & Y & -- & -- & -- \\
LMC-14426 & $6.384(7)$ & $220.8(17)$ & 7855.13528 & 7905.19163 & V1 & 1865 & 37/55 & 82/110 & 5/5/5 & 8/8/8 & Y & -- & -- & -- \\
LMC-14846 & $13.6514(8)$ & $378.4(20)$ & 10002.66570 & 9974.29418 & V2 & 5780 & 81/113 & 134/181 & 6/6/6 & 10/10/10 & Y & -- & -- & -- \\
LMC-16854 & $4.32781(4)$ & $169.9(15)$ & 9995.71186 & 9989.68002 & V2 & 5797 & 146/250 & 218/334 & 16/16/16 & 19/19/19 & Y & -- & -- & -- \\
LMC-17632 & $3.982750(13)$ & $152.29(15)$ & 6313.45517 & 6385.60705 & V1 & 6235 & 230/416 & 299/495 & 19/19/19 & 20/19/20 & Y & Y & -- & -- \\
LMC-18374 & $4.2855550(29)$ & $125.00(21)$ & 7856.61848 & 7843.80182 & V1 & 5690 & 148/253 & 230/358 & 16/16/16 & 20/20/20 & Y & Y & Y & -- \\
LMC-20092 & $8.138257(22)$ & $154.36(6)$ & 9959.97587 & 9899.65902 & V2 & 6947 & 267/378 & 304/462 & 19/19/19 & 20/20/20 & Y & Y & Y & -- \\
LMC-21665 & $5.116994(8)$ & $171.17(16)$ & 5688.40667 & 5671.98609 & V2 & 1301 & 23/32 & 160/266 & 1/1/1 & 7/7/7 & -- & -- & Y & -- \\
LMC-22012 & $7.613002(12)$ & $267.32(31)$ & 5593.18086 & 5598.93172 & V1 & 1381 & 41/66 & 179/287 & 2/2/2 & 7/7/7 & -- & -- & -- & -- \\
LMC-22288 & $8.644970(9)$ & $344.8(7)$ & 9972.61840 & 9976.48588 & V2 & 6473 & 230/346 & 269/416 & 10/10/10 & 10/10/10 & Y & -- & -- & -- \\
LMC-23767 & $7.626966(9)$ & $273.53(11)$ & 6325.51972 & 6262.37471 & V1 & 1511 & 107/174 & 187/293 & 5/5/5 & 8/8/8 & Y & Y & Y & Y \\
LMC-30826 & $6.514800(23)$ & $225.68(8)$ & 5456.22111 & 5465.78801 & V1 & 6680 & 256/414 & 340/507 & 19/19/19 & 20/20/20 & Y & Y & Y & Y \\
LMC-30850 & $10.74744(29)$ & $181.40(16)$ & 9983.20398 & 9929.99170 & V2 & 4938 & 153/288 & 247/349 & 17/14/17 & 18/18/18 & Y & -- & Y & -- \\
LMC-32315 & $6.571730(5)$ & $188.3^{\mathrm{u}}$ & 9960.37500 & 9908.38081 & V2 & 7005 & 249/392 & 353/510 & 19/19/19 & 20/20/20 & Y & Y & Y & Y \\
LMC-33694 & $6.93423(7)$ & $242.67(7)$ & 9982.27104 & 10043.40141 & V2 & 6258 & 210/296 & 233/385 & 17/17/16 & 19/17/19 & Y & Y & -- & -- \\
LMC-35460 & $7.75192(6)$ & $302.08(5)$ & 6309.16703 & 6392.20979 & V1 & 1341 & 20/39 & 64/250 & 1/1/1 & 7/3/7 & -- & -- & Y & Y \\
LMC-36460 & $10.33462(21)$ & $413.6(8)$ & 6321.24556 & 6511.63716 & V1 & 1529 & 40/59 & 206/295 & 2/2/2 & 9/9/9 & Y & Y & Y & -- \\
LMC-36665 & $6.63321(5)$ & $185.40(21)$ & 5593.49621 & 5595.52137 & V1 & 6381 & 233/357 & 306/478 & 18/18/18 & 20/20/20 & Y & Y & -- & -- \\
LMC-37483 & $5.01318(9)$ & $172.53(11)$ & 5594.59970 & 5668.01127 & V1 & 1773 & 39/170 & 254/486 & 5/2/5 & 10/9/10 & Y & -- & -- & -- \\
SMC-01144 & $4.13666^{\mathrm{u}}$ & $135.48(13)$ & 6270.72611 & 6237.86930 & V1 & 3061 & 162/349 & 280/491 & 9/8/9 & 10/10/10 & Y & -- & -- & -- \\
SMC-01807 & $4.84152^{\mathrm{u}}$ & $166.38(6)$ & 6272.00590 & 6290.43632 & V1 & 3202 & 193/362 & 243/422 & 9/9/9 & 9/9/9 & Y & Y & Y & Y \\
SMC-01830 & $6.835805(5)$ & $266.1(11)$ & 6272.24869 & 6315.19449 & V1 & 3201 & 194/308 & 278/446 & 9/9/9 & 10/10/10 & Y & Y & Y & -- \\
SMC-01983 & $6.37710^{\mathrm{u}}$ & $228.7^{\mathrm{u}}$ & 9873.70530 & 9958.97454 & V2 & 3078 & 182/313 & 276/444 & 9/9/9 & 10/10/10 & Y & -- & -- & -- \\
SMC-02049 & $7.115545(11)$ & $182.0(5)$ & 9870.39025 & 9807.56029 & V2 & 3247 & 206/326 & 274/418 & 9/9/9 & 10/10/10 & Y & -- & -- & -- \\
SMC-02090 & $3.462641(4)$ & $106.352(21)$ & 9874.00786 & 9837.76775 & V2 & 3073 & 108/183 & 193/352 & 5/5/5 & 7/7/7 & -- & -- & Y & Y \\
SMC-04933 & $6.10998^{\mathrm{u}}$ & $202.01(10)$ & 9878.94045 & 9868.64970 & V2 & 3083 & 154/251 & 243/410 & 7/7/7 & 10/10/10 & Y & -- & Y & -- \\
SMC-05046 & $4.192315(7)$ & $137.85(5)$ & 6292.31424 & 6252.67568 & V1 & 2793 & 110/208 & 223/390 & 6/6/6 & 10/10/10 & Y & Y & Y & -- \\
SMC-05062 & $4.782388(5)$ & $148.820(19)$ & 6281.84665 & 6240.91665 & V1 & 2927 & 172/327 & 250/431 & 9/9/9 & 10/10/10 & Y & -- & -- & -- \\
SMC-05637 & $5.6021574(12)$ & $149.97(16)$ & 9881.77910 & 9903.86139 & V2 & 3000 & 188/318 & 244/402 & 9/9/9 & 10/10/10 & Y & -- & -- & -- \\
SMC-06532 & $9.9939(6)$ & $185.5(9)$ & 10139.44622 & 10198.55948 & V2 & 2374 & 56/86 & 79/190 & 6/5/6 & 9/7/9 & Y & -- & -- & -- \\
SMC-09290 & $4.99531(8)$ & $163.87(10)$ & 9917.35345 & 9972.99732 & V2 & 2487 & 53/86 & 157/252 & 4/4/4 & 9/9/9 & Y & Y & Y & -- \\
SMC-09371 & $7.03266(9)$ & $239.77(18)$ & 9921.38817 & 9829.36133 & V2 & 2482 & 69/96 & 129/196 & 5/5/5 & 8/8/8 & Y & Y & -- & -- \\
SMC-09439 & $4.539750(13)$ & $178.06(19)$ & 7901.28858 & 7851.10116 & V1 & 2363 & 89/149 & 124/206 & 6/6/6 & 8/8/8 & Y & -- & -- & -- \\
SMC-09816 & $5.99805(5)$ & $194.05(7)$ & 7901.78998 & 7899.28980 & V1 & 2489 & 47/63 & 150/246 & 3/3/3 & 9/8/9 & Y & -- & -- & -- \\
SMC-10065 & $5.11471^{\mathrm{u}}$ & $125.6^{\mathrm{u}}$ & 7903.03107 & 7858.22636 & V1 & 2273 & 66/111 & 161/227 & 5/5/5 & 10/10/10 & Y & -- & -- & -- \\
SMC-10197 & $4.555903(5)$ & $168.62(10)$ & 10164.70906 & 10208.22091 & V2 & 2355 & 101/180 & 137/218 & 8/8/8 & 9/9/9 & Y & -- & Y & -- \\
SMC-10426 & $4.771472(23)$ & $156.80(5)$ & 7900.16856 & 7879.44363 & V1 & 2344 & 57/107 & 120/185 & 6/6/6 & 8/8/8 & Y & Y & Y & -- \\
\end{longtable}
\noindent{Notes.} Epochs are HJD minus 2450000 (days): fixed phase origins, not individually timed events; their tabulation precision is not an uncertainty. Identifiers abbreviate OGLE-LMC/SMC-ECL-number as cloud-number. Periods are in days; parentheses give conditional one-standard-error uncertainties in the last digits, and superscript $u$ marks an unstable local error. E marks timing evaluability; Tim. marks $Q$/$T$ modulation; Prof. marks the broad indicator, including displacement-like responses. $O$/$W$ marks residual wing or width variations (Appendix~\ref{app:sensitivity}). These selections share data and are not statistically independent. V1 and V2 denote the curation branches in Table~\ref{tab:referee_selection_flow}.
\endgroup

\clearpage

Table~\ref{tab:referee_nonretained54} supplies an explicit present-day timing-evaluability reason for each of the 54 candidates outside the curated sample. The working periods in that table support this retrospective check and are not replacements for the periods of the 55-system inference. In particular, a candidate with an uncertain long clock or too few adequately covered long-phase intervals cannot be counted among systems with no detected eclipse-center change.

\begingroup\footnotesize\renewcommand{\arraystretch}{0.9}\setlength{\tabcolsep}{4pt}
\begin{longtable}{@{}lrrrrc@{}}
\caption{Candidates not retained: present suitability for long-phase eclipse timing}\label{tab:referee_nonretained54}\\
\hline\hline
Object & $P_{\rm orb}$ & $P_{\rm long}$ & $C_0$ & $C_{0.5}$ & Reason \\
\hline
\endfirsthead
\caption[]{continued.}\\
\hline\hline
Object & $P_{\rm orb}$ & $P_{\rm long}$ & $C_0$ & $C_{0.5}$ & Reason \\
\hline
\endhead
\hline
\multicolumn{6}{r}{continued on next page}\\
\endfoot
\hline\hline
\endlastfoot
OGLE-LMC-ECL-00750 & 4.97336 & 172.4 & 0 & 0 & C \\
OGLE-LMC-ECL-01600 & 1.69970 & 46.5 & 0 & 0 & C \\
OGLE-LMC-ECL-02547 & 7.75880 & -- & -- & -- & L \\
OGLE-LMC-ECL-04277 & 20.98878 & 810.2 & 5 & 6 & C \\
OGLE-LMC-ECL-06671 & 4.88511 & 159.3 & 0 & 3 & C \\
OGLE-LMC-ECL-08494 & 5.71315 & 177.4 & 2 & 3 & C \\
OGLE-LMC-ECL-09132 & 63.75505 & 1498.2 & 1 & 3 & C \\
OGLE-LMC-ECL-10831 & 28.49880 & 916.5 & 0 & 4 & C \\
OGLE-LMC-ECL-15202 & 3.46482 & 339.6 & 0 & 0 & C \\
OGLE-LMC-ECL-25090 & 5.20575 & 496.4 & 0 & 1 & C \\
OGLE-LMC-ECL-25186 & 4.12711 & 123.6 & 0 & 0 & C \\
OGLE-LMC-ECL-26412 & 3.38617 & 182.5 & 0 & 0 & C \\
OGLE-LMC-ECL-26818 & 11.14709 & 186.1 & 1 & 3 & C \\
OGLE-LMC-ECL-27222 & 6.71512 & 221.9 & 0 & 0 & C \\
OGLE-LMC-ECL-27688 & 6.19666 & 178.5 & 0 & 2 & C \\
OGLE-LMC-ECL-28145 & 4.39408 & 142.6 & 0 & 0 & C \\
OGLE-LMC-ECL-30441 & 9.12921 & 376.9 & 4 & 7 & R \\
OGLE-LMC-ECL-31070 & 4.84943 & 296.2 & 0 & 0 & C \\
OGLE-LMC-ECL-31289 & 8.30749 & 312.2 & 0 & 2 & C \\
OGLE-LMC-ECL-31364 & 4.85121 & 159.2 & 0 & 0 & C \\
OGLE-LMC-ECL-32186 & 6.03928 & 191.1 & 0 & 0 & C \\
OGLE-LMC-ECL-34095 & 4.88603 & 123.6 & 0 & 1 & C \\
OGLE-LMC-ECL-34137 & 5.61344 & 206.285 & 0 & 2 & C \\
OGLE-LMC-ECL-35461 & 5.62073 & 188.762 & 4 & 6 & C \\
OGLE-LMC-ECL-37220 & 5.62931 & 174.7 & 1 & 3 & C \\
OGLE-LMC-ECL-37476 & 5.80378 & 188.8 & 0 & 5 & C \\
OGLE-LMC-ECL-38058 & 5.52212 & 160.2 & 0 & 1 & C \\
OGLE-LMC-ECL-38592 & 12.75807 & 209.1 & 0 & 3 & C \\
OGLE-LMC-ECL-39428 & 89.33549 & 626.1 & 1 & 2 & C \\
OGLE-LMC-ECL-39629 & 5.39859 & 185.5 & 0 & 0 & C \\
OGLE-LMC-ECL-45119 & 13.81458 & 576.7 & 0 & 0 & C \\
OGLE-LMC-ECL-45455 & 5.34492 & -- & -- & -- & L \\
OGLE-LMC-ECL-47760 & 8.01512 & 490.6 & 0 & 1 & C \\
OGLE-LMC-ECL-51664 & 4.32093 & 138.0 & 0 & 0 & C \\
OGLE-LMC-ECL-51858 & 4.00467 & 274.4 & 0 & 0 & C \\
OGLE-LMC-ECL-52047 & 7.11777 & 221.6 & 0 & 0 & C \\
OGLE-LMC-ECL-53750 & 9.21400 & 345.0 & 0 & 2 & C \\
OGLE-SMC-ECL-03555 & 17.64267 & 569.1 & 4 & 6 & C \\
OGLE-SMC-ECL-05481 & 8.75766 & -- & -- & -- & L \\
OGLE-SMC-ECL-09133 & 5.17743 & 169.3 & 3 & 4 & C \\
OGLE-SMC-ECL-09570 & 0.45689 & -- & -- & -- & L \\
OGLE-SMC-ECL-1003 & 5.10733 & 420.1 & 0 & 0 & C \\
OGLE-SMC-ECL-1005 & 4.45508 & 152.6 & 0 & 0 & C \\
OGLE-SMC-ECL-10251 & 7.48847 & 289.8 & 4 & 6 & C \\
OGLE-SMC-ECL-1028 & 4.15961 & 129.2 & 0 & 0 & C \\
OGLE-SMC-ECL-1402 & 7.47956 & 286.1 & 0 & 0 & C \\
OGLE-SMC-ECL-1799 & 7.15595 & 223.6 & 0 & 0 & C \\
OGLE-SMC-ECL-2331 & 4.91589 & 160.0 & 0 & 0 & C \\
OGLE-SMC-ECL-2519 & 3.98769 & -- & -- & -- & L \\
OGLE-SMC-ECL-2597 & 28.58473 & 1337.9 & 0 & 0 & C \\
OGLE-SMC-ECL-3248 & 4.71303 & 131.3 & 0 & 0 & C \\
OGLE-SMC-ECL-3604 & 5.02386 & 409.0 & 1 & 0 & C \\
OGLE-SMC-ECL-4181 & 5.05441 & 176.9 & 1 & 0 & C \\
OGLE-SMC-ECL-4756 & 0.62853 & 34.1 & 0 & 0 & C \\
\end{longtable}
\noindent{Notes.} Periods are working values (d) for this retrospective check, not new adopted periods for the 55-system inference. $C_0$ and $C_{0.5}$ count coverage-valid long-phase bins. Reason L: no secure long-period clock; C: neither eclipse attains the required long-phase coverage; R: a single marginal eclipse reaches seven bins but does not satisfy the robustness requirement. For OGLE-LMC-ECL-02547 the long period remains ambiguous; the visually selected diagnostic periods of OGLE-LMC-ECL-34137 and 35461 are 206.285 and 188.762 d, respectively. These reasons describe current evaluability, not the full historical reason for each visual-curation decision, and no row is a timing non-detection.
\endgroup

\clearpage

Table~\ref{tab:app_colours} gives the 27 objects with at least one locally interpolated eclipse color. It distinguishes pair counts from groups sharing I endpoints, corrected colors and the reddening-independent contrast. Sightline map uncertainty is shared by both eclipse colors and does not contribute independently to their difference.

\begingroup\footnotesize\setlength{\tabcolsep}{3pt}
\begingroup\footnotesize\renewcommand{\arraystretch}{0.90}\setlength{\tabcolsep}{3pt}
\begin{longtable}{@{}llrlrlr@{}}
\caption{Eclipse colors using local $I$-band interpolation}\label{tab:app_colours}\\
\hline\hline
Object & $E(V-I)$ & $N_0/G_0$ & $C^{\rm map}_{\rm dered,0}$ & $N_{0.5}/G_{0.5}$ & $C^{\rm map}_{\rm dered,0.5}$ & $\Delta C_{0-0.5}$ \\
\hline
\endfirsthead
\caption[]{continued.}\\
\hline\hline
Object & $E(V-I)$ & $N_0/G_0$ & $C^{\rm map}_{\rm dered,0}$ & $N_{0.5}/G_{0.5}$ & $C^{\rm map}_{\rm dered,0.5}$ & $\Delta C_{0-0.5}$ \\
\hline
\endhead
\hline\endfoot
\hline\hline\endlastfoot
OGLE-LMC-ECL-05532 & $0.115^{+0.057}_{-0.054}$ & 5/5 & $0.145\pm0.033$ & 0/0 & -- & -- \\
OGLE-LMC-ECL-05935 & $0.099^{+0.058}_{-0.063}$ & 1/1 & $0.033\pm0.007$ & 0/0 & -- & -- \\
OGLE-LMC-ECL-06957 & $0.094^{+0.056}_{-0.056}$ & 2/2 & $0.257\pm0.022$ & 2/2 & $0.104\pm0.009$ & $0.153\pm0.024$ \\
OGLE-LMC-ECL-07098 & $0.088^{+0.035}_{-0.042}$ & 3/3 & $0.929\pm0.026$ & 4/4 & $0.351\pm0.019$ & $0.578\pm0.032$ \\
OGLE-LMC-ECL-11994 & $0.123^{+0.123}_{-0.094}$ & 1/1 & $0.093\pm0.008$ & 2/2 & $0.031\pm0.007$ & $0.062\pm0.010$ \\
OGLE-LMC-ECL-12848 & $0.073^{+0.104}_{-0.094}$ & 0/0 & -- & 2/2 & $-0.029\pm0.014$ & -- \\
OGLE-LMC-ECL-14202 & $0.122^{+0.059}_{-0.058}$ & 1/1 & $1.248\pm0.035$ & 3/3 & $0.487\pm0.043$ & $0.761\pm0.056$ \\
OGLE-LMC-ECL-14426 & $0.052^{+0.060}_{-0.059}$ & 0/0 & -- & 1/1 & $0.538\pm0.034$ & -- \\
OGLE-LMC-ECL-18374 & $0.085^{+0.049}_{-0.058}$ & 2/2 & $0.319\pm0.016$ & 1/1 & $0.020\pm0.010$ & $0.299\pm0.019$ \\
OGLE-LMC-ECL-23767 & $0.203^{+0.091}_{-0.084}$ & 1/1 & $0.476\pm0.033$ & 0/0 & -- & -- \\
OGLE-LMC-ECL-30826 & $0.126^{+0.103}_{-0.085}$ & 3/3 & $0.323\pm0.036$ & 2/2 & $0.170\pm0.028$ & $0.153\pm0.046$ \\
OGLE-LMC-ECL-36665 & $0.289^{+0.169}_{-0.134}$ & 0/0 & -- & 3/3 & $0.102\pm0.010$ & -- \\
OGLE-SMC-ECL-10065 & $0.042^{+0.027}_{-0.030}$ & 0/0 & -- & 1/1 & $0.140\pm0.015$ & -- \\
OGLE-LMC-ECL-07343 & $0.119^{+0.075}_{-0.056}$ & 2/2 & $1.085\pm0.020$ & 2/2 & $0.693\pm0.024$ & $0.391\pm0.031$ \\
OGLE-LMC-ECL-09173 & $0.099^{+0.075}_{-0.061}$ & 1/1 & $-0.072\pm0.012$ & 3/3 & $-0.114\pm0.010$ & $0.041\pm0.016$ \\
OGLE-LMC-ECL-11330 & $0.067^{+0.084}_{-0.081}$ & 3/3 & $0.022\pm0.032$ & 3/3 & $-0.073\pm0.012$ & $0.095\pm0.035$ \\
OGLE-LMC-ECL-11906 & $0.104^{+0.107}_{-0.087}$ & 3/3 & $0.531\pm0.008$ & 0/0 & -- & -- \\
OGLE-LMC-ECL-11961 & $0.097^{+0.071}_{-0.054}$ & 3/3 & $0.752\pm0.084$ & 2/2 & $0.470\pm0.034$ & $0.282\pm0.090$ \\
OGLE-LMC-ECL-12704 & $0.096^{+0.042}_{-0.056}$ & 2/2 & $0.186\pm0.008$ & 1/1 & $0.150\pm0.006$ & $0.036\pm0.010$ \\
OGLE-LMC-ECL-12808 & $0.086^{+0.107}_{-0.098}$ & 3/3 & $0.504\pm0.011$ & 0/0 & -- & -- \\
OGLE-LMC-ECL-13898 & $0.074^{+0.054}_{-0.051}$ & 0/0 & -- & 1/1 & $0.044\pm0.012$ & -- \\
OGLE-LMC-ECL-14413 & $0.106^{+0.086}_{-0.079}$ & 3/3 & $0.957\pm0.021$ & 1/1 & $1.193\pm0.012$ & $-0.236\pm0.024$ \\
OGLE-LMC-ECL-14846 & $0.068^{+0.078}_{-0.070}$ & 4/4 & $0.391\pm0.025$ & 0/0 & -- & -- \\
OGLE-LMC-ECL-16854 & $0.077^{+0.051}_{-0.055}$ & 1/1 & $0.157\pm0.018$ & 2/2 & $0.114\pm0.021$ & $0.043\pm0.028$ \\
OGLE-LMC-ECL-20092 & $0.101^{+0.072}_{-0.075}$ & 2/2 & $0.580\pm0.045$ & 1/1 & $0.038\pm0.009$ & $0.542\pm0.046$ \\
OGLE-LMC-ECL-22288 & $0.286^{+0.140}_{-0.117}$ & 2/2 & $0.678\pm0.061$ & 3/3 & $-0.017\pm0.027$ & $0.695\pm0.067$ \\
OGLE-LMC-ECL-33694 & $0.083^{+0.056}_{-0.061}$ & 3/3 & $0.093\pm0.028$ & 1/1 & $0.130\pm0.010$ & $-0.036\pm0.030$ \\
\end{longtable}
\noindent{Notes.} Colors and errors are in magnitudes. Each $I$ endpoint lies within 0.01 orbital cycles of its $V$ epoch, which itself lies within 0.01 cycles of the calculated eclipse center. $N$ counts pairs and $G$ counts groups connected by a shared $I$ endpoint. Cluster resampling and shared $I$-error draws account for that dependence; errors remain conditional on the adopted periods, interpolation and sampled epochs. Map dispersions in the second column are shared sightline systematics, not formal independent color errors. Template-curvature differences are assessed separately and not treated as exact corrections. A missing region is not assigned a color.
\endgroup

\endgroup

\section{Sensitivity of the timing procedure}\label{app:sensitivity}

The profile-shape analysis distinguishes a displacement-like change from changes in the eclipse wings and width. Here I specify the four fixed shapes fitted to each long-phase profile. Let $s$ be the nominal eclipse half-width. Subtracting the median magnitude in the outer wings of the reference profile ($|x|\geq0.8s$) gives the mean eclipse shape $B(x)$. On its fixed phase grid, the derivative $D(x)=dB/dx$ describes the first-order effect of a small translation. Two further trial shapes are $O_0(x)=(x/s)^3B(x)/\max|B|$, sensitive mainly to unequal wings, and $W_0(x)=(x/s)D(x)$, sensitive mainly to width.

I first fitted $O_0$ with the best-fitting combination of a constant level, $B$, and $D$ on that grid; after subtracting this combination and dividing the remainder by its maximum absolute value, I call the result $O(x)$. I then treated $W_0$ in the same way, this time subtracting the best-fitting combination of a constant, $B$, $D$, and the already defined $O$; normalization of this second remainder gives $W(x)$. These preparatory fits minimize the unweighted sum of squared differences on the common phase grid. They remove overlap among the {trial shapes}, not components from an observed eclipse or physically separate sources of light.

The observed magnitude profile is subsequently fitted as $c+\alpha B+\gamma_DD+\gamma_OO+\gamma_WW$. Because the reference eclipse may be asymmetric, $O$ and $W$ need not have exact odd or even parity. Their fitted coefficients describe profile changes; none is an independent time of stellar conjunction.

For reproducibility, the Q fit uses inverse propagated flux-variance weights, iteratively reducing residuals beyond 1.345 robust-scatter units. Each wing requires five points and the vertex remains within the nominal half-width of the calculated center. T requires ten points, at least three on each side, a positive scale between 0.1 and 3, and a shift within half its fitted support; its magnitude residuals use the same robust cutoff. Harmonic fits require at least seven centers. Detection additionally requires phase coverage $1-|\langle e^{2\pi i\phL}\rangle|\geq0.25$ and an uncertainty-inflation factor no larger than three, comparing leave-interval-out amplitude scatter with bootstrap amplitude scatter. Excess scatter is estimated from residuals and included in harmonic weights; the 2000 coefficient-pair draws retain this scatter and the center errors. These quality checks precede the significance selection and are not additional FDR tests.

Table~\ref{tab:referee_sensitivity} shows how the detections change when I vary the local $Q$ window, replace its parabola by a cubic, or change the $T$ reference profile. Only one center-measurement procedure changes in each row. To make the rows comparable, the multiple-testing correction retains the original list of candidate curves, even when a curve cannot be measured by an alternative method. The table also gives the number of curves that remain measurable. In the independent-template rows, each reference and target uses only half the data; many $T$ curves then lack enough long-phase coverage. Their smaller detection counts chiefly reflect this loss of data, not evidence that fewer DPVs have variable eclipses.

\begin{table}[H]\caption{Sensitivity of timing selections to the measurement procedure.}\label{tab:referee_sensitivity}\centering\small\begin{tabular}{@{}lrrrrr@{}}\hline\hline Procedure & Systems & Curves & Fitted $Q$/$T$ & Admissible $Q$/$T$ & Lost/gained \\\hline
Baseline & 27 & 46 & 89/90 & 79/81 & -- \\
Local cubic $Q$ & 19 & 35 & 82/90 & 73/81 & 9/1 \\
$T$ omits measured interval & 26 & 40 & 89/80 & 79/72 & 1/0 \\
$Q$ window $0.75s$ & 19 & 36 & 66/90 & 57/81 & 8/0 \\
$Q$ window $1.00s$, no fallback & 25 & 43 & 82/90 & 74/81 & 2/0 \\
$Q$ window $1.25s$ & 25 & 46 & 86/90 & 75/81 & 2/0 \\
$T$ reference A, target B & 23 & 29 & 89/17 & 79/14 & 4/0 \\
$T$ reference B, target A & 23 & 29 & 89/20 & 79/15 & 4/0 \\
\hline\end{tabular}\tablefoot{Fitted means that the harmonic can be estimated; admissible additionally requires the detection-quality criteria. Lost/gained counts systems relative to the baseline. Each row changes only one method; the other stays basal. The original hypothesis list is retained and an unmeasurable alternative is assigned $p=1$. In the split-template rows, reduced data volume prevents many fits. $s$ is the nominal eclipse half-width. Of nine systems lost with the cubic, seven retain an admissible formerly selected curve but no longer pass FDR; two fail admissibility. These are method checks, not alternative population incidences.}\end{table}

\begin{figure}[H]
\centering
\includegraphics[width=\textwidth]{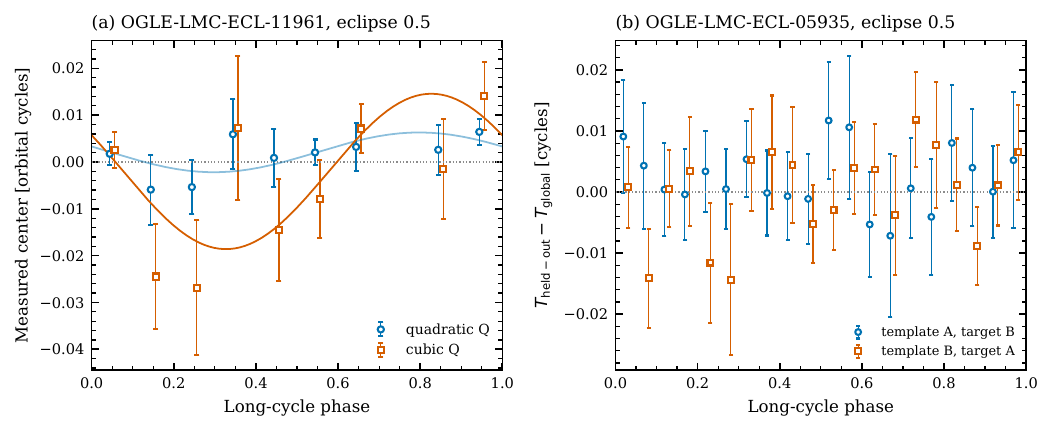}
\caption{Examples of how the measured eclipse center depends on the fitting choice. Left: $Q$ centers of the phase-0.5 eclipse of OGLE-LMC-ECL-11961 obtained with a local parabola or a cubic; the two fitted long-cycle patterns need not agree. Right: for the phase-0.5 eclipse of OGLE-LMC-ECL-05935, the change in $T$ when the same target measurements are fitted with an independently constructed reference profile instead of the global reference. The two symbol types show the two ways of dividing the observations into template and target halves. Vertical bars show the scatter of the paired bootstrap differences and are not formal confidence intervals.}
\label{fig:sensitivity_examples}
\end{figure}

Figure~\ref{fig:sensitivity_examples} illustrates why the table is necessary. In OGLE-LMC-ECL-11961, changing the local $Q$ function can alter a selected eclipse-center pattern. In OGLE-LMC-ECL-05935, measuring the same target points against the global and independent-half $T$ references isolates the influence of template construction. These examples do not establish a preferred function or a universal $T$ systematic error.

\clearpage
Table~\ref{tab:template_objects} summarizes the paired reference comparisons for all 55 objects, pooling the two assignments of independent template and target halves. Its medians and percentiles are descriptive sensitivity scales. A small signed average across the sample does not remove object-dependent offsets, and these scales have not been added to baseline errors or propagated to baseline FDR selections.
\begingroup\footnotesize\renewcommand{\arraystretch}{0.90}\setlength{\tabcolsep}{3pt}
\begin{longtable}{@{}lrrrrrr@{}}
\caption{Sensitivity of $T$ to reference construction for each curated system}\label{tab:template_objects}\\
\hline\hline
Object & $N_0$ & $M_0$ & $P_{90,0}$ & $N_{0.5}$ & $M_{0.5}$ & $P_{90,0.5}$ \\
\hline
\endfirsthead
\caption[]{continued.}\\
\hline\hline
Object & $N_0$ & $M_0$ & $P_{90,0}$ & $N_{0.5}$ & $M_{0.5}$ & $P_{90,0.5}$ \\
\hline
\endhead
\hline\endfoot
\hline\hline\endlastfoot
OGLE-LMC-ECL-00778 & 6 & 1.06 & 2.84 & 16 & 1.45 & 4.69 \\
OGLE-LMC-ECL-04725 & 14 & 4.00 & 16.61 & 13 & 5.81 & 25.15 \\
OGLE-LMC-ECL-05532 & 39 & 3.73 & 17.21 & 40 & 3.82 & 11.83 \\
OGLE-LMC-ECL-05935 & 38 & 2.17 & 5.61 & 40 & 4.20 & 11.57 \\
OGLE-LMC-ECL-06376 & 4 & 1.40 & 3.73 & 13 & 2.91 & 10.88 \\
OGLE-LMC-ECL-06957 & 32 & 2.25 & 6.54 & 39 & 6.79 & 14.79 \\
OGLE-LMC-ECL-07098 & 18 & 0.51 & 1.89 & 18 & 1.31 & 3.50 \\
OGLE-LMC-ECL-07343 & 20 & 1.14 & 2.71 & 20 & 1.85 & 8.69 \\
OGLE-LMC-ECL-09173 & 32 & 0.33 & 0.60 & 38 & 0.44 & 1.48 \\
OGLE-LMC-ECL-11330 & 35 & 2.46 & 7.37 & 37 & 2.75 & 10.48 \\
OGLE-LMC-ECL-11906 & 20 & 0.87 & 2.57 & 20 & 1.74 & 8.14 \\
OGLE-LMC-ECL-11961 & 16 & 2.81 & 5.38 & 18 & 1.88 & 9.36 \\
OGLE-LMC-ECL-11994 & 38 & 0.46 & 3.42 & 40 & 2.17 & 7.88 \\
OGLE-LMC-ECL-12704 & 14 & 1.79 & 4.99 & 20 & 4.55 & 11.73 \\
OGLE-LMC-ECL-12808 & 20 & 2.05 & 4.65 & 19 & 3.33 & 19.72 \\
OGLE-LMC-ECL-12848 & 32 & 1.75 & 6.91 & 38 & 3.51 & 10.41 \\
OGLE-LMC-ECL-13898 & 35 & 0.60 & 1.16 & 40 & 2.87 & 10.49 \\
OGLE-LMC-ECL-14202 & 12 & 0.78 & 1.77 & 16 & 3.62 & 6.54 \\
OGLE-LMC-ECL-14413 & 16 & 0.75 & 2.59 & 20 & 1.08 & 2.37 \\
OGLE-LMC-ECL-14426 & 10 & 2.44 & 4.58 & 16 & 4.11 & 15.22 \\
OGLE-LMC-ECL-14846 & 10 & 1.61 & 9.59 & 18 & 11.87 & 24.04 \\
OGLE-LMC-ECL-16854 & 31 & 1.11 & 2.64 & 38 & 2.14 & 11.05 \\
OGLE-LMC-ECL-17632 & 38 & 1.62 & 6.00 & 40 & 1.60 & 10.36 \\
OGLE-LMC-ECL-18374 & 30 & 0.74 & 2.31 & 38 & 3.58 & 8.99 \\
OGLE-LMC-ECL-20092 & 38 & 0.16 & 0.52 & 40 & 2.02 & 5.93 \\
OGLE-LMC-ECL-21665 & 2 & 2.40 & 3.52 & 14 & 0.89 & 3.87 \\
OGLE-LMC-ECL-22012 & 4 & 0.79 & 1.45 & 14 & 2.15 & 8.64 \\
OGLE-LMC-ECL-22288 & 20 & 0.31 & 0.68 & 20 & 1.15 & 4.97 \\
OGLE-LMC-ECL-23767 & 10 & 1.00 & 2.15 & 16 & 2.52 & 12.77 \\
OGLE-LMC-ECL-30826 & 38 & 1.84 & 6.76 & 40 & 2.32 & 14.42 \\
OGLE-LMC-ECL-30850 & 32 & 3.20 & 22.77 & 34 & 3.38 & 17.60 \\
OGLE-LMC-ECL-32315 & 38 & 1.70 & 9.37 & 40 & 2.46 & 12.26 \\
OGLE-LMC-ECL-33694 & 30 & 6.74 & 18.51 & 38 & 7.82 & 31.95 \\
OGLE-LMC-ECL-35460 & 2 & 17.53 & 24.84 & 12 & 15.46 & 31.07 \\
OGLE-LMC-ECL-36460 & 4 & 3.37 & 41.64 & 16 & 2.95 & 14.23 \\
OGLE-LMC-ECL-36665 & 33 & 3.09 & 12.56 & 40 & 6.83 & 20.67 \\
OGLE-LMC-ECL-37483 & 10 & 1.87 & 15.29 & 19 & 4.00 & 10.88 \\
OGLE-SMC-ECL-01144 & 18 & 2.18 & 12.98 & 18 & 2.85 & 13.35 \\
OGLE-SMC-ECL-01807 & 18 & 1.20 & 2.29 & 18 & 2.11 & 8.28 \\
OGLE-SMC-ECL-01830 & 18 & 1.32 & 4.15 & 20 & 3.94 & 11.09 \\
OGLE-SMC-ECL-01983 & 18 & 1.28 & 3.00 & 20 & 1.42 & 8.50 \\
OGLE-SMC-ECL-02049 & 17 & 0.76 & 2.51 & 20 & 2.49 & 4.57 \\
OGLE-SMC-ECL-02090 & 10 & 0.94 & 3.75 & 14 & 6.64 & 9.33 \\
OGLE-SMC-ECL-04933 & 14 & 0.68 & 1.52 & 20 & 1.83 & 4.18 \\
OGLE-SMC-ECL-05046 & 12 & 0.77 & 1.33 & 20 & 2.64 & 9.81 \\
OGLE-SMC-ECL-05062 & 18 & 0.54 & 1.69 & 20 & 2.10 & 4.22 \\
OGLE-SMC-ECL-05637 & 18 & 1.35 & 2.01 & 20 & 0.94 & 2.26 \\
OGLE-SMC-ECL-06532 & 11 & 0.99 & 20.36 & 17 & 1.64 & 22.42 \\
OGLE-SMC-ECL-09290 & 8 & 4.58 & 14.07 & 18 & 1.56 & 10.71 \\
OGLE-SMC-ECL-09371 & 10 & 2.37 & 15.85 & 16 & 7.12 & 39.04 \\
OGLE-SMC-ECL-09439 & 12 & 1.71 & 6.32 & 15 & 1.87 & 6.52 \\
OGLE-SMC-ECL-09816 & 5 & 2.29 & 9.36 & 18 & 1.41 & 9.35 \\
OGLE-SMC-ECL-10065 & 9 & 2.41 & 7.78 & 20 & 2.76 & 15.62 \\
OGLE-SMC-ECL-10197 & 16 & 3.22 & 12.14 & 17 & 4.36 & 19.32 \\
OGLE-SMC-ECL-10426 & 12 & 1.81 & 6.80 & 16 & 2.97 & 10.14 \\
\end{longtable}
\noindent{Notes.} $M$ and $P_{90}$ are the median and 90th percentile of $|T_{\rm independent}-T_{\rm global}|$, in $10^{-3}$ orbital cycles, fitting identical target points. $N$ counts usable interval--half comparisons, pooling both template/target assignments; they are not independent eclipse events. These are sensitivity scales, not systematic standard errors or confidence limits, and are not propagated into the baseline FDR selection. Individual bootstrap differences can be non-regular or approach fit bounds; the paired products retain those cautions.
\endgroup

\clearpage
\section{Observed O$-$C examples}\label{app:examples}

\begin{figure}[H]
\sidecaption
\includegraphics[width=12cm]{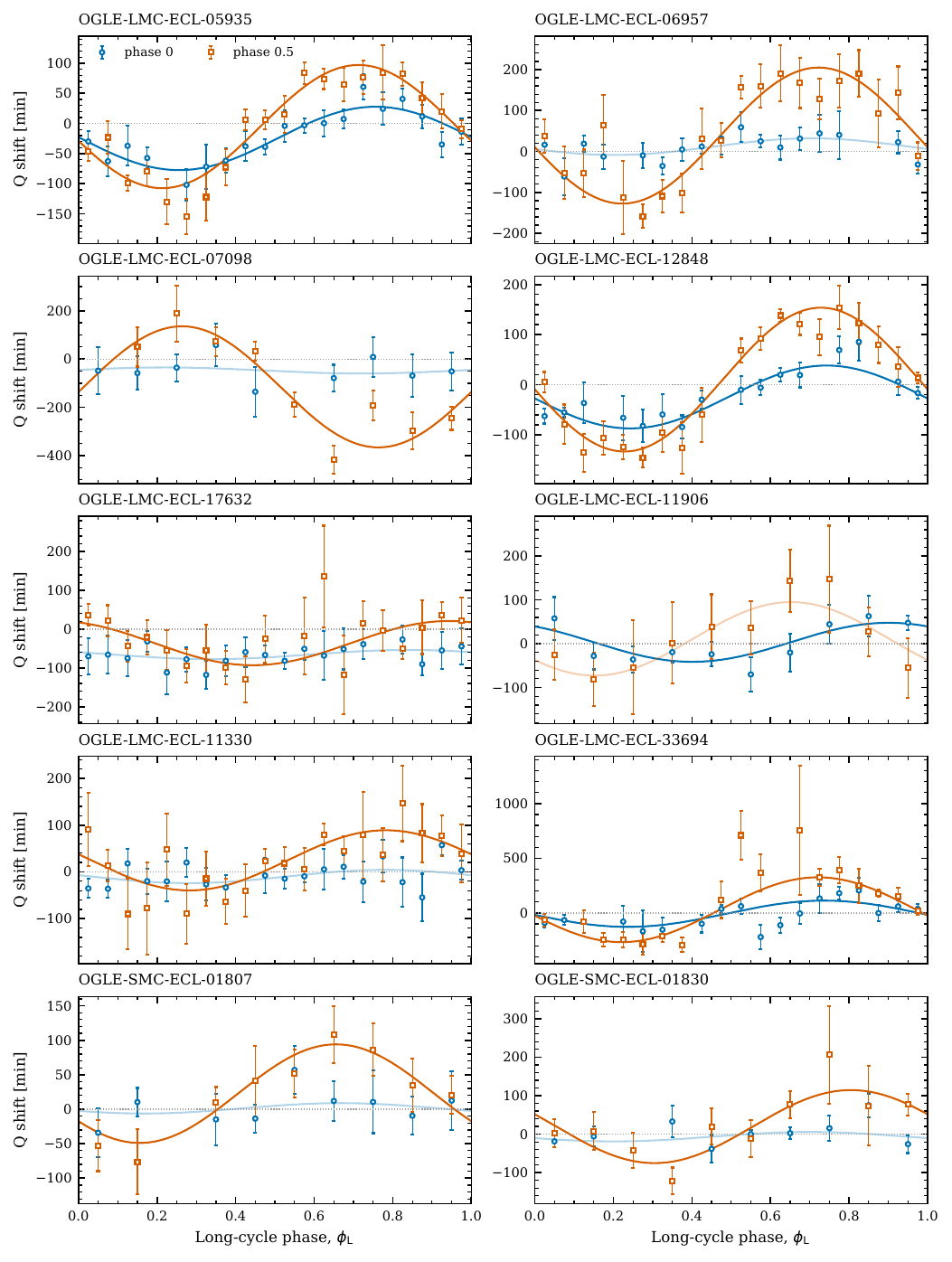}
\caption{Ten illustrative observed $Q$ eclipse-center curves from LMC and SMC DPVs. Blue circles and orange squares correspond to the operational eclipses at phases 0 and 0.5. Points show centers from non-overlapping long-phase profiles and conjunction-level bootstrap standard errors; curves are revised first-harmonic fits, faint when not individually selected. The conversion to minutes uses each object's adopted orbital period.}
\label{fig:ten_oc}
\end{figure}

Figure~\ref{fig:ten_oc} shows ten representative observed $Q$ curves, including LMC and SMC objects, using the revised conjunction-level errors. They illustrate that a selected long-phase modulation can differ between the two eclipses; the display is not a selection-free atlas. Each displacement in minutes uses the orbital period of that particular system. The curves describe photometric centers, not directly measured conjunction times.

\clearpage
\section{Further diagnostic of the artificial system}\label{app:model}

\begin{figure}[H]
\sidecaption
\includegraphics[width=12cm]{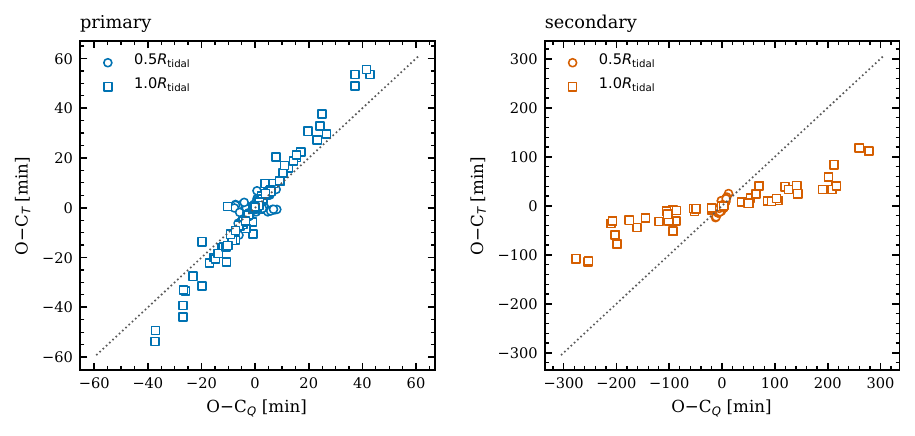}
\caption{$Q$ versus $T$ for the synthetic long-phase configurations at the two operational eclipses. The dotted line is equality. Circles and squares denote $R_{\rm d}=0.5\Rt$ and $R_{\rm d}=\Rt$, respectively; both joint rim-region states are included. Large departures from equality arise in asymmetric noise-free profiles, particularly for the large-disk phase-0.5 eclipse.}
\label{fig:sim_qt}
\end{figure}

Figure~\ref{fig:sim_qt} compares the two center definitions for every synthetic eclipse profile. Their separation, especially in large-disk phase-0.5 eclipses, arises from the same noise-free brightness distribution and is therefore not a result of OGLE measurement scatter. The comparison makes explicit why $Q$ and $T$ are complementary descriptions of an eclipse rather than interchangeable estimates of an orbital event.

\end{appendix}

\end{document}